\documentclass[twocolumn, journal]{IEEEtran}

\usepackage{amsmath}
\usepackage{amssymb}
\usepackage{bm}
\usepackage{graphics}
\usepackage{graphicx}
\usepackage{booktabs}
\usepackage{subcaption}
\usepackage{cite}

\newcommand\D{\ensuremath{\bm{D}}}

\newcommand\lp{\ensuremath{\left(}}
\newcommand\rp{\ensuremath{\right)}}

\newcommand\Reals{\ensuremath{\mathbb{R}}}

\newcommand\GP[1]{\ensuremath{ {\lp #1 \rp }}}

\newcommand\GB[1]{\ensuremath{ {\left[ #1 \right] }}}
\newcommand\GN[1]{\ensuremath{ {\left|\left| #1 \right|\right| }}}
\newcommand\GM[1]{\ensuremath{ {\left| #1 \right| }}}

\newcommand\GC[1]{\ensuremath{ {\left\{ #1 \right\} }}}
\newcommand\argmin{\ensuremath{\mathop{\text{argmin }}}}

\renewcommand\min{\ensuremath{\mathop{\text{min}}}}

\newcommand\grad{\ensuremath{\nabla}}
\newcommand\hess{\ensuremath{\nabla^2}}
\newcommand\diag{\ensuremath{\mathop{\text{diag}}}}

\newcommand\eref[1]{(\ref{#1})}

\newcommand\btheta{\ensuremath{\bm \theta}}

\newcommand\eg{{e.g.}, }

\newcommand\ie{{i.e.}, }

\newcommand\R{\ensuremath{\mathsf{R}}}

\newcommand\LL{\ensuremath{\mathsf{L}}}

\newcommand\A{\ensuremath{{\bm{A}}}}
\newcommand\bE{\ensuremath{{\bm{E}}}}

\newcommand\x{\ensuremath{{\bm{x}}}}

\newcommand\f{\ensuremath{{\bm{f}}}}

\newcommand\y{\ensuremath{{\bm{y}}}}

\newcommand\W{\ensuremath{\bm{W}}}

\newcommand\G{\ensuremath{\bm{G}}}
\newcommand\tr{\ensuremath{^\intercal}}

\newcommand\M{\ensuremath{{\bm{M}}}}

\newcommand\X{\ensuremath{{\bm{X}}}}

\newcommand\w{\ensuremath{{\bm{w}}}}
\newcommand\g{\ensuremath{{\bm{g}}}}

\newcommand\z{\ensuremath{{\bm{z}}}}
\renewcommand\r{\ensuremath{{\bm{r}}}}

\newcommand\bT{\ensuremath{\bm{T}}}
\newcommand\bB{\ensuremath{\bm{B}}}
\newcommand\bX{\ensuremath{\bm{X}}}
\newcommand\bR{\ensuremath{\bm{R}}}
\newcommand\btau{\ensuremath{\bm{\tau}}}
\newcommand\bM{\ensuremath{\bm{M}}}
\newcommand\bC{\ensuremath{\bm{C}}}
\newcommand\bS{\ensuremath{\bm{S}}}
\newcommand\bTheta{\ensuremath{{\bm{\Theta}}}}

\newcommand\bD{\ensuremath{{\bm{D}}}}

\newcommand\ones{\ensuremath{{\bm{1}}}}
\newcommand\zeros{\ensuremath{{\bm{0}}}}

\newcommand\I{\ensuremath{\bm{I}}}

\renewcommand\w{\ensuremath{{\bm{w}}}}
\newcommand\p{\ensuremath{{\bm{p}}}}

\newcommand\sn{\ensuremath{^{(n)}}}
\newcommand\snp{\ensuremath{^{(n+1)}}}

\newcommand\snz{\ensuremath{^{(0)}}}

\title{A practical light transport system model
for chemiluminescence distribution reconstruction}

\author{Madison G. McGaffin,
\textit{Member, IEEE,}
\quad
Hao Chen,
\quad
Jeffrey A. Fessler
\textit{Fellow, IEEE,}
\quad
Volker Sick
\thanks{%
Supported in part by National Science Foundation grant
CBET 1402707.
Any opinion, findings, and conclusions or recommendations expressed
in this material are those of the authors
and do not necessarily reflect the views of the
National Science Foundation.

The original version of this paper
was submitted to IEEE Trans. on Computational Imaging
on 2016-11-17,
right after the end of the first author's postdoctoral appointment.
Detailed reviews came back on 2017-03-10
requesting various revisions,
but the first author has moved to industry
and no longer works in this area.
There were no fundamental errors
mentioned by the reviewers.
The biggest requests were for comparisons with related work.
The following citations
were also suggested
\cite{%
dansereau:13:dca,
klehm:13:vst,
todoroff:14:roi,
wetzstein:10:ssi,
ihrke:04:ibt,
schwarz:96:mtf,
li:14:vio,
hasinoff:03:pc3%
}.

M.~McGaffin was and J.~Fessler is
with the EECS Department,
and H.~Chen and V.~Sick
are with the ME Department
of the University of Michigan.
emails:
\texttt{\{
fessler,haochen,vsick\}@umich.edu }
}
}

\begin{document}

\maketitle

\begin{abstract}
Plenoptic cameras and other integral photography instruments capture
richer angular information from a scene than traditional 2D cameras.
This extra information is used to estimate depth,
perform superresolution or reconstruct 3D information from the scene.
Many of these applications involve
solving a large-scale numerical optimization problem.
Most published approaches model the camera(s)
using pre-computed matrices that require large amounts of memory
and are not well-suited to modern many-core processors.
We propose a flexible camera model based on light transport
and use it to model plenoptic and traditional cameras.
We implement the proposed model on a GPU
and use it to reconstruct simulated
and real 3D chemiluminescence distributions (flames)
from images taken by traditional and plenoptic cameras.
\end{abstract}

\section{Introduction}
\label{sec,intro}

Plenoptic cameras~\cite{ng:05:lfp,perwass:12:sl3}, cameras with coded
masks~\cite{marwah:13:clf}, and other integral photography equipment augment
traditional 2D photographic measurements with additional angular information.
They do this by interposing additional lenses, masks or other elements along
the optical path from the scene to the detector; these additional provide
angular information about the scene that is ``integrated away'' by traditional
cameras.  The spatial/angular data acquired by one of these devices can
facilitate depth estimation, digital refocusing, superresolution, and, in the
main application of this paper, tomographic reconstruction of chemilumiscence
distributions, \ie flames.

The 3D structure of translucent luminescent objects is relevant for multiple
mechanical engineering and modeling
tasks~\cite{elsinga:06:tpi,fahringer:12:tro,greene:13:vrf}.  Previous works
have used multiple traditional or plenoptic cameras to acquire enough angular
information to reconstruct the object~\cite{shroff:13:ifa,cai:13:pao}.  This is
often done by numerically solving an inverse problem, and a key part of that
problem is the model that predicts from a candidate 3D chemiluminescence
distribution, $\x$, the resulting image on a camera, $\A\x$.  Although these
works have taken different approaches to modeling the physics of the
acquisition process, the most common practice has been to precompute the
(sparse but large) matrix $\A$ and use sparse linear algebra routines to solve
the reconstruction problem.  While this technique can produce good results, it
is computationally expensive and solving even relatively small problems can
take many hours even using GPU linear algebra libraries, in part because
computing speed gains have outpaced memory bandwidth increases in modern
many-core computing systems.

This work proposes a practical light transport-based framework
for the camera model, $\A$.
Instead of precomputing the system matrix $\A$,
we provide expressions to compute its entries on the fly,
which is significantly more efficient.
We then describe an efficient GPU implementation.
The system is ``light transport-based''
in the sense that it numerically implements
a discrete version of certain analytical light transport techniques%
~\cite{liang:11:lfa,liang:15:alt} to analyze camera properties.
This structure results in camera models
that are compositions of simple light transport steps;
this enables modeling a wide range of camera designs
with only a few tools.
We present results quantifying the accuracy of the proposed system model
and apply it to several image reconstruction problems
with simulated and real data.
Mathematically, the chemiluminescence reconstruction problem
is similar to SPECT image reconstruction%
~\cite{tsui:88:ios,leahy:00:sai};
in both cases the goal is to
reconstruct the spatial distribution of the rate of emission of photons
using models for the system physics.
Unlike traditional SPECT,
the chemiluminescence reconstruction problem we consider
gathers information only from a few fixed camera positions,
and therefore has relatively sparse angular information about the object.

Section~\ref{sec,model} describes how we implement the light transport
``building block'' operations.
Section~\ref{sec,chemi} models 3D chemiluminescence
distributions.  Section~\ref{sec,practical} discusses some implementation
practicalities, Section~\ref{sec,recon} proposes a multi-camera reconstruction
algorithm for a chemiluminescence distribution, and Section~\ref{sec,experiment}
contains some experiments.  Section~\ref{sec,conc} contains some concluding
remarks.


\section{Camera system model}
\label{sec,model}

This section derives computational models for the progression of light from a
scene to the camera's detector.  The main tool we use is light transport via
geometric optics.  The model describes how an incident light field propagates
stage-by-stage through the camera to the detector.  To accomplish this, we
discretize (\ie define a finite-series representation of) the light field at
each stage spatially (\eg on the detector along pixel boundaries) and angularly
(by where the described light field passes through a designated ``angular
plane.'').  The resulting model is highly parallelizable and has
computationally efficient forward and transpose operations (those operations
are essential for use in many inverse problem settings).

\subsection{Geometric optics}
\label{sec,optics}

\begin{figure}
    \centering
    \includegraphics[width=1.5in]{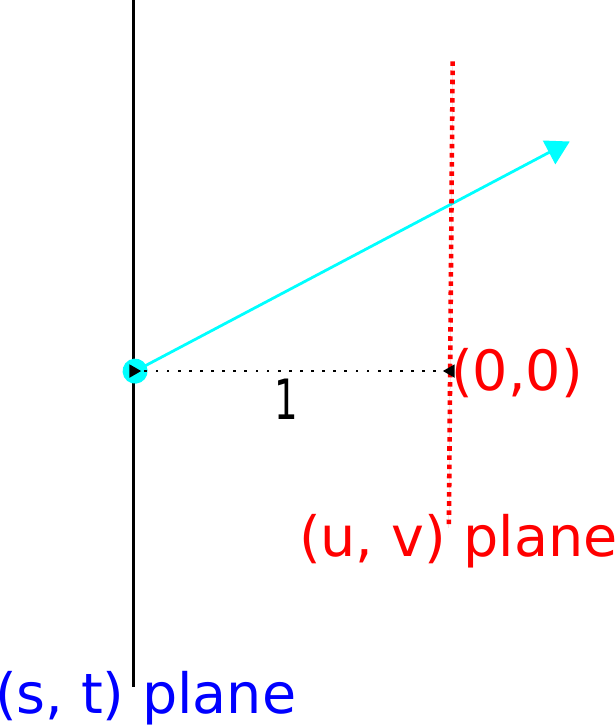}
    \caption{Two-plane ray parameterization}
    \label{fig,two,plane}
\end{figure}

\begin{figure}
    \centering
    \includegraphics[width=3in]{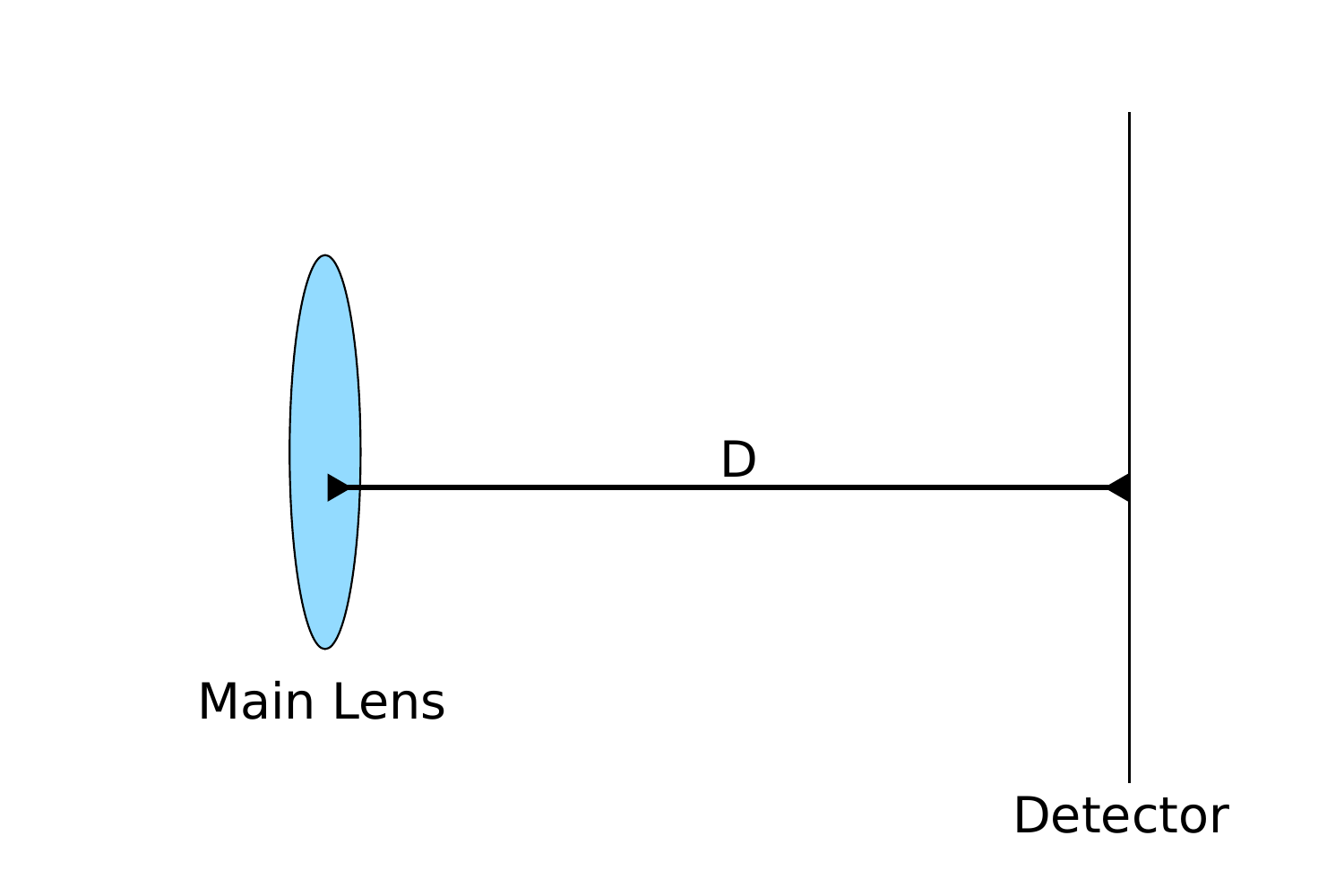}
    \caption{Schematic of a single-lens camera}
    \label{fig,single}
\end{figure}

In this paper we model light transport using geometric optics, following
earlier works, \eg \cite{liang:15:alt,liang:11:lfa}.  Because we are interested
in modeling the behavior of cameras imaging scenes at macro scale, we do not
need to resort to wave transport \cite{broxton:13:wot} to model the imaging
process.  This section summarizes the elements of light transport used
in the proposed system model.

The monochrome light field function defined at a given plane at a fixed time is
a four dimensional real-valued function $\LL: \Reals^4 \to \Reals$.  The light
field function has two spatial arguments $\GP{s, t}$ and two angular arguments
$\GP{u, v}$, using the common two-plane parameterization~\cite{levoy:96:lfr};
see Figure~\ref{fig,two,plane}.  Together these coordinates $\btheta = \GP{s,
u, t, v}$ describe the position and angular orientation of a ray passing
through the light field's $(s, t)$ plane.  The corresponding value of the light field
function at that point gives the radiance along that ray.

\begin{table}
    \centering
    \caption{Common affine optical transformations}
    \label{tab,optics}
    \begin{tabular}{ll}
        \toprule
        {\bf Operation} & {\bf Transformation} \\
        \midrule
        Propagation by $d$ & 
        $
        \bT_d\GP\btheta
        =
        \GB{%
            \begin{matrix}
                1 & d & 0 & 0 \\
                0 & 1 & 0 & 0 \\
                0 & 0 & 1 & d \\
                0 & 0 & 0 & 1 
            \end{matrix}
        }
        \GB{%
            \begin{matrix}
                s \\ u \\ t \\ v
            \end{matrix}
        }
        $
        \\
        \begin{minipage}{1in}
            Thin lens refraction \\
            with focal length $f$ \\
            and center \\
            ${\GP{s, t}=\GP{s_0, t_0}}$
        \end{minipage}
        &
        $
        \bR_f\GP\btheta
        =
        \GB{%
            \begin{matrix}
                1 & 0 & 0 & 0 \\
                -1/f & 1 & 0 & 0 \\
                0 & 0 & 1 & 0 \\
                0 & 0 & -1/f & 1 
            \end{matrix}
        }
        \GB{%
            \begin{matrix}
                s \\ u \\ t \\ v
            \end{matrix}
        }
        $
        \\
        &
        $
        \hspace{15em}+
        \GB{%
            \begin{matrix}
                0 \\ s_0/f \\ 0 \\ t_0/f
            \end{matrix}
        }
        $
        \\
        \bottomrule
    \end{tabular}
\end{table}

Geometric optics describes how rays of light are altered, \ie how the
parameters $\btheta$ change, as rays pass through space and refracting media;
including, particularly for this paper, ideal thin lenses and occluding masks.
Ignoring diffraction and attenuation, light transport corresponds to
transformations of $\btheta$ that are affine and easy to compute.
Table~\ref{tab,optics} gives expressions for the transformations used in this
paper.

The expressions in Table~\ref{tab,optics} are composed to describe more complex
optical effects.  For example, a ray $\btheta$ entering the camera in
Figure~\ref{fig,single} is refracted by the main lens, then travels the distance $D$
before landing on the detector.  That is, the light field on the detector,
$\LL^\text{det}\GP\btheta$, is determined by the light field impinging on the
main lens, $\LL^\text{main}\GP\btheta$:
\begin{align}
    \LL^\text{main}\GP\btheta &= \LL^\text{det}\GP{\X^\text{det-main}\GP{\btheta}},
\end{align}
where
$
\X^\text{det-main}\GP\btheta
=
 \GP{\bT_D \circ \bR_{f_\text{main}}}\GP{\btheta}
,$
and $\circ$ denotes function composition.
For any two planes $p$ and $q$ in an optical system,
we use the notation $\bX^{pq}$ to describe the optical transformation
from $q$ to $p$.

Because we will refer to specific entries of such optical
transformations later in this paper, we denote any such the
optical transformation $\bX$ as
\begin{align}
    \GB{\begin{matrix}
            X_{\mathrm s}\GP{\btheta} \\
            X_{\mathrm u}\GP{\btheta} \\
            X_{\mathrm t}\GP{\btheta} \\
            X_{\mathrm v}\GP{\btheta}
    \end{matrix}}
    &=
    \GB{\begin{matrix}
            X_\mathrm{ss} & X_\mathrm{su} & 0 & 0 \\
            X_\mathrm{us} & X_\mathrm{uu} & 0 & 0 \\
            0 & 0 & X_\mathrm{tt} & X_\mathrm{tv} \\
            0 & 0 & X_\mathrm{vt} & X_\mathrm{vv} \\
    \end{matrix}}
    \GB{\begin{matrix}
            s \\ u \\ t \\ v
    \end{matrix}}
    +
    \GB{\begin{matrix}
        \overline X_\mathrm{s} \\ \overline X_\mathrm{u} \\ \overline X_\mathrm{t} \\ \overline X_\mathrm{v}
    \end{matrix}}
    .
\end{align}
The separable structure follows from the ideal geometric optics in
Table~\ref{tab,optics} and helps yield convenient computational structures.  It
is possible to extend the work in this paper to handle non-separable optical
transformations, and one could use the transformation factorization approach
described in Section~\ref{sec,rot} to implement many non-separable
transformations efficiently.

\subsection{Light field discretization}
\label{sec,discretization}

Consider a plane normal to the primary axis of the optical system.
To store and compute the light field at that plane,
we approximate it with a basis expansion,
as is common in other inverse problems%
~\cite{censor:83:fse,lewitt:03:oom}.
Light field
$\LL\GP{\btheta}$ has two spatial and two angular variables;
we discretize the spatial dimension $(s,t)$
using separable pillbox functions, \ie pixels.
To handle the angular coordinates $(u,v)$,
we designate a plane in the optical system
and discretize the light field
by where the ray $\btheta$ lands on that plane.
Often, but not always, this plane is along the main lens of the camera.
We call this plane the ``angular plane'' of the system.
In our approach,
all light fields in the optical system use the same angular plane
for angular discretization.

Let $\X^{0p}$ denote the optical transformation
from the plane $p$
(including all lenses and free-space propagation)
to the selected angular plane.
We represent the light field at the plane $p$
using the following basis expansion with coefficients
$\f^p \in \Reals^{N^p K}$:
\begin{align}
    \LL^p\GP{\btheta; \f^p}
    &=
    \sum_{k=1}^K
    a\GP{\frac{X^{0p}_\mathrm{s}\GP{s,u} - s_k}{\Delta_{\mathrm s}^0}}
    a\GP{\frac{X^{0p}_\mathrm{t}\GP{t,v} - t_k}{\Delta_{\mathrm t}^0}}
    \nonumber \\
    &\times
    \sum_{i=1}^{N^p}
    b\GP{\frac{s - s_i}{\Delta_{\mathrm s}^p}}
    b\GP{\frac{t - t_i}{\Delta_{\mathrm t}^p}}
    f^p_{ki}
    \label{eqn,lf,disc}
,\end{align}
where $N^p$ denotes the number of pixels
and $K$ denotes the number of angular coordinate samples
(sub-aperture images).
The basis center points $\GC{\GP{s_k, t_k}}$ on the angular plane and,
(lightly recycling notation)
$\GC{\GP{s_i, t_i}}$ on the plane $p$,
are separated by the respective distances
$\GC{\GP{\Delta_{\mathrm s}^0, \Delta_{\mathrm t}^0}}$
and
$\GC{\GP{\Delta_{\mathrm s}^p, \Delta_{\mathrm t}^p}}$,
respectively.
The spatial basis function $b\GP{\cdot}$
is the standard rectangular function:
\begin{align}
    b\GP{t}
    &=
    \begin{cases}
        1, & \GM{t} \le \frac{1}{2} \\
        0, & \text{else.}
    \end{cases}
\end{align}
We consider two choices for the basis function on the angular plane,
$a\GP{\cdot}$:
the rect function $b$,
that leads to a 2D ``pillbox'' basis,
and the Dirac impulse $\delta$.
Using
$a = \delta$ yields slightly simpler expressions in Section~\ref{sec,xport}
for light transport,
but we found that it often requires a finer
(and therefore more computationally expensive) discretization
of the angular plane to produce an accurate model.
The Dirac impulse basis is used implicitly
when a finite camera lens is modeled
as a superposition of pinhole cameras~\cite{nien:14:phd}.
Section~\ref{sec,angular}
explores this trade-off.

\subsection{Light transport}
\label{sec,xport}

Let $p$ and $q$ be two planes in the optical system with the same optical
plane.  Assume that $q$ is closer to the scene and $p$ is closer to the
detector; to model the camera's image acquisition process, we model the
transport from $q$ to $p$.

The optical transformations from $q$ and $p$ to the angular plane are
$\bX^{0q}$ and $\bX^{0p}$, respectively.  From these expressions, the optical
transformation from $q$ to $p$
is $\btheta^p = \bX^{pq}\GP{\btheta^q}$ where
$\bX^{pq} = {\GB{\bX^{0p}}^{-1}\circ\bX^{0q}}$.
In general, when we start with a light field having the representation~\eref{eqn,lf,disc},
after a transformation it will no longer have exactly that same representation.
This property is acceptable since~\eref{eqn,lf,disc} is already an approximation of the
continuous light field.  To maintain~\eref{eqn,lf,disc} as a consistent form of the
representation throughout the model, after each optical transformation we
project the transformed light field onto a finite dimensional subspace
of the form~\eref{eqn,lf,disc}.  Specifically,
to
find the coefficients
of the discrete light field on $p$, $\f^p$, from the coefficients of the
light field on $q$, $\f^q$, we solve the following optimization problem
in $L_2\GP{\Reals^4}$:
\begin{align}
    \f^p
    &=
    \argmin_{\f}
    \GN{\LL^p\GP{\bX^{pq}\GP\cdot; \f} - \LL^q\GP{\cdot; \f^q}}_2^2.
    \label{eqn,xport}
\end{align}
This least-squares approximation problem has a block-separable solution
with $K$ blocks, $\f^p = \GP{\f^p_1, \ldots, \f^p_K}$; one block for each
basis function on the angular plane:
\begin{align}
    \f^p_k = \frac{1}{V^p} \bB^{pq}_k \f^q_k,
\end{align}
where the ``volume'' of a basis element in $\Reals^4$ on plane $p$ is
defined by
\begin{align}
    &V^p
    =
    \GN{%
        a\GP{\frac{X^{0p}_{\mathrm s}\GP{s,u}}{\Delta_{\mathrm s}^0}}
        a\GP{\frac{X^{0p}_{\mathrm t}\GP{t,v}}{\Delta_{\mathrm t}^0}}
        b\GP{\frac{s}{\Delta_{\mathrm s}^p}}
        b\GP{\frac{t}{\Delta_{\mathrm t}^p}}
    }_2^2.
\end{align}
The entries of $\bB^{pq}_k$ come from the $\GP{s, t, u, v}$
inner products:
\begin{align}
    \GB{\bB^{pq}_k}_{ij}
    &=
    \left<
        b\GP{\frac{X^{pq}_{\mathrm s}\GP{s,u} - s_i}{\Delta_{\mathrm s}^p}}
        a\GP{\frac{X^{0q}_{\mathrm s}\GP{s,u} - s_k}{\Delta_{\mathrm s}^0}}
        ,
        \right.
    \nonumber \\
    & \left. \hspace{4em}
        b\GP{\frac{s - s_j}{\Delta_{\mathrm s}^q}}
    \right>
    \times
    \nonumber \\
    &\hspace{1.1em}
    \left<
        b\GP{\frac{X^{pq}_{\mathrm t}\GP{t,v} - t_i}{\Delta_{\mathrm t}^p}}
        a\GP{\frac{X^{0q}_{\mathrm t}\GP{t,v} - t_k}{\Delta_{\mathrm t}^0}}
        ,
        \right.
    \nonumber \\
    & \left. \hspace{4em}
        b\GP{\frac{t - t_j}{\Delta_{\mathrm t}^q}}
    \right>,
    \label{eqn,xport,ip}
    \intertext{%
        where the first inner product is over $\GP{s,u}$ and the
        second is over $\GP{t, v}$.
        After some simplification,}
    \GB{\bB^{pq}_k}_{ij}
    &=
    \int_{s_j-\Delta_\mathrm{s}/2}^{s_j + \Delta_\mathrm{s}/2}
        h_{k, \mathrm s}^{pq} g\GP{s - \alpha^{pq}_{k, \mathrm s} s_i; \btau^{pq}_{k, \mathrm s}}~\text{d}s
    \nonumber \\
    &\hspace{3em}
    \times
    \int_{t_j-\Delta_\mathrm{t}/2}^{t_j + \Delta_\mathrm{t}/2}
        h_{k, \mathrm t}^{pq} g\GP{t - \alpha^{pq}_{k, \mathrm t} t_i; \btau^{pq}_{k, \mathrm t}}~\text{d}t.
    \label{eqn,xport,int}
\end{align}
The blur kernel $g$ results from an inner integral over $u$ or $v$ and
depends on the choice of angular basis function $a$.
The magnification terms $\GC{\alpha^{pq}_{k,*}}$ and blur parameters
$\GC{\btau^{pq}_{k,*}}$ depend only on the planes $p$ and $q$ and the angle $k$.
For the Dirac $\delta$ and pillbox angular basis functions, the blur integrals
and blur parameters are efficient to derive and compute.  Tables~\ref{tab,dirac}
and~\ref{tab,pillbox} give expressions for these parameters for the
$s$ direction; the $t$ direction expressions are analogous.

\begin{table}
    \centering
    \caption{Dirac basis transport expressions}
    \label{tab,dirac}
    \begin{tabular}{ll}
        \toprule
        {\bf Property} & {\bf Expression} \\
        \midrule
        Blur kernel
        &
        $g\GP{s} = \begin{cases}
            1, & \tau_{s1}^{pqk} \le w < \tau_{s2}^{pqk} \\
            0, &\text{else.}
        \end{cases}$
        \\
        &
        $\alpha = X^{pq}_\mathrm{ss} - X^{0q}_\mathrm{ss} X^{pq}_\mathrm{su} / X^{0q}_\mathrm{su} $
        \\
        &
        $\beta = \overline X^{pq}_\mathrm{s} + X^{pq}_\mathrm{su}\GP{s_k - \overline X^{0q}_\mathrm{s}}/X^{0q}_\mathrm{su}$
        \\
        &
        $\widehat \btau_s^{pqk} = \GP{\pm \Delta_\mathrm{s}^p/2 - \beta}/\alpha$
        \\
        Blur parameters
        &
        $\btau_s^{pqk} = \text{sort-ascend}\GC{\widehat \btau_s^{pqk}}$
        \\
        Blur height
        &
        $ 
        h^{pqk}_s = \Delta^0_s / \GM{X^{0q}_\mathrm{su}}
        $
        \\
        Blur magnification
        &
        $\alpha^{pqk}_s = 1 / \alpha$
        \\
        \midrule
        Basis element volume
        &
        $V^p = \Delta_{\mathrm s}^p \Delta_{\mathrm t}^p 
            \GM{\Delta_{\mathrm s}^0 \Delta_{\mathrm t}^0 / \GP{X^{0p}_\mathrm{su} X^{0p}_\mathrm{tv}} }$
        \\
        \bottomrule
    \end{tabular}
\end{table}

\begin{table}
    \centering
    \caption{Pillbox basis transport expressions}
    \label{tab,pillbox}
    \begin{tabular}{ll}
        \toprule
        {\bf Property} & {\bf Expression} \\
        \midrule
        Blur kernel
        &
        $g\GP{s}
        =
        \begin{cases}
            \frac{s - \tau_{s1}^{pqk}}{\tau_{s2}^{pqk} - \tau_{s1}^{pqk}},
                & \tau_{s1}^{pqk} \le s < \tau_{s2}^{pqk}; \\
            1, & \tau_{s2}^{pqk} \le s < \tau_{s3}^{pqk}; \\
            1 - \frac{s - \tau_{s3}^{pqk}}{\tau_{s4}^{pqk} - \tau_{s3}^{pqk}}
                & \tau_{s3}^{pqk} \le s < \tau_{s4}^{pqk}; \\
            0 & \text{else.}
        \end{cases}
        $
        \\
        &
        $\alpha = X^{pq}_\mathrm{ss} - X^{pq}_\mathrm{su} X^{0q}_\mathrm{ss} / X^{0q}_\mathrm{su} $
        \\
        &
        $\beta = X^{pq}_\mathrm{su} / X^{0q}_\mathrm{su} $
        \\
        &
        $\gamma = \overline X^{pq}_{\mathrm s} - X^{pq}_\mathrm{su} \overline X^{0q}_{\mathrm s} / X^{0q}_\mathrm{su}$
        \\
        &
        $ \widehat \btau_s^{pqk} = \GP{\pm \Delta_{\mathrm s}^p/2 - \beta\GP{s_k \mp \Delta_{\mathrm s}^0/2}
            - \gamma}/\alpha$
            \\
        Blur parameters
        &
        $ \btau_s^{pqk} = \text{sort-ascend}\GC{ \widehat \btau_s^{pqk} }$
        \\
        Blur height
        &
        $ \min\GC{\GM{\Delta_{\mathrm s}^0 / X^{0q}_\mathrm{su}}, \GM{\Delta_{\mathrm s}^p / X^{pqk}_\mathrm{su}}} $
        \\
        Blur magnification
        &
        $ \alpha_s^{pqk} = 1 / \alpha $
        \\
        \midrule
        &
        $m_{\mathrm s} = \text{max}\GC{\Delta_{\mathrm s}^0 / 2 \GM{X^{0p}_\mathrm{su}}, 
            \Delta_{\mathrm s}^p \GM{X^{0p}_\mathrm{ss} / X^{0p}_\mathrm{su}} / 2}$
        \\
        &
        $h_s = \text{min}\GC{\Delta_{\mathrm s}^p, \Delta_{\mathrm s}^0 / \GM{X^{0p}_\mathrm{ss}}}$
        \\
        &
        $V^p_s = 2 m_s h_s$
        \\
        Basis element volume
        &
        $V^p = V^p_s V^p_t$
        \\
        \bottomrule
    \end{tabular}
\end{table}

Note that the entries of $\bB^{pq}_k$ are separable products of one-dimensional
$s$ and $t$ functions~\eref{eqn,xport,int}.  Consequently, we implement
$\bB^{pq}_k$ as the Kronecker product
\begin{align}
    \bB^{pq}_k = \bB^{pq}_{k \mathrm s} \otimes \bB^{pq}_{k \mathrm t}.
    \label{eqn,xport,sep}
\end{align}

\subsection{Occlusion}
\label{sec,occlusion}

The final important optical elements we need to model are occluders, \eg a coded
mask inside the camera~\cite{marwah:13:clf} or outside the
lens~\cite{babacan:12:clf}.  For an occluder on the plane $p$, let
$\LL^{p-}$ be the light field at $p$ on the side of the occluder towards the
scene and $\LL^{p+}$ be the light field at $p$ on the side towards the
detector.  We model occlusion in terms of the light field coefficients as
\begin{align}
    \f_k^{p+} &= \bM \f_k^{p-},
    \label{eqn,occlusion}
\end{align}
where $\bM$ is a $N^p \times N^p$ diagonal matrix that encodes the occluder's
spatial behavior.  We compute the entries of $\bM$ by rasterizing the
occluder's support onto the spatial grid for $\LL^p$.

\subsection{Measurement formation}
\label{sec,signal}

Let $\LL^d$ be the light field on the detector, and we choose the spatial
discretization of $\LL^d$ to align with the spatial discretization of the
detector.  In the absence of noise, the $i$th sensor measurement is the
integral of the irradiance of the light over the spatial extent of the $i$th
sensor cell, and with this choice of discretization the integral simplifies as
follows:
\begin{align}
    y_i
    &=
    \int_{s \in S_i}\int_{t \in T_i}\int_{u \in \Reals}\int_{v \in \Reals}
    \LL^d\GP{s, u, t, v}
    ~\text{d}s
    ~\text{d}u
    ~\text{d}t
    ~\text{d}v
    \nonumber \\
    &=
    \sqrt{V^d}
    \sum_{k=1}^K
    f^d_{ki}.
\end{align}
Thus the vector of measurements $\y$ is simply the sum over each angular
component
\begin{align}
    \y &= \sqrt{V_d} \sum_{k=1}^K \f_k^d.
\end{align}

\subsection{Camera models}
\label{sec,camera}

We combine the operations in the preceding sections to model single-lens and
plenoptic cameras.  For both types of cameras, we place the angular plane on the
main lens, and consider transporting a scene light field $\LL^s$, parameterized
by $\f^s$ located $D_\text{scene}$ units from the camera's main lens to
the detector plane.  The
optical transformation from $\LL^s$ to the angular plane is $\bX^{0s} =
\GP{\bT_{D_\text{scene}} \circ \bR_{f_\text{main}}}^{-1}$, where $\bR_{f\text{main}}$
denotes refraction by the main lens.

\subsubsection{Single lens camera}

Modeling a single lens camera,
Figure~\ref{fig,single},
requires a single transport operation,
from the scene light field $\LL^s$
through the main lens
to the detector light field $\LL^d$.
The optical transformation
from the detector to the angular plane
is the propagation $\bX^{0d} = \bT_{D}$,
where $D$ is the distance from the detector to the main lens.
With $\bX^{0d}$ and $\bX^{0s}$ defined,
we can follow Sections~\ref{sec,xport} and~\ref{sec,signal}
to write the single lens camera system model:
\begin{align}
    \y_\text{single}
    &=
    \sqrt{V^d} \sum_{k=1} \bB^{ds}_k \f^s_k.
\end{align}

\begin{figure}
    \centering
    \includegraphics[width=3in]{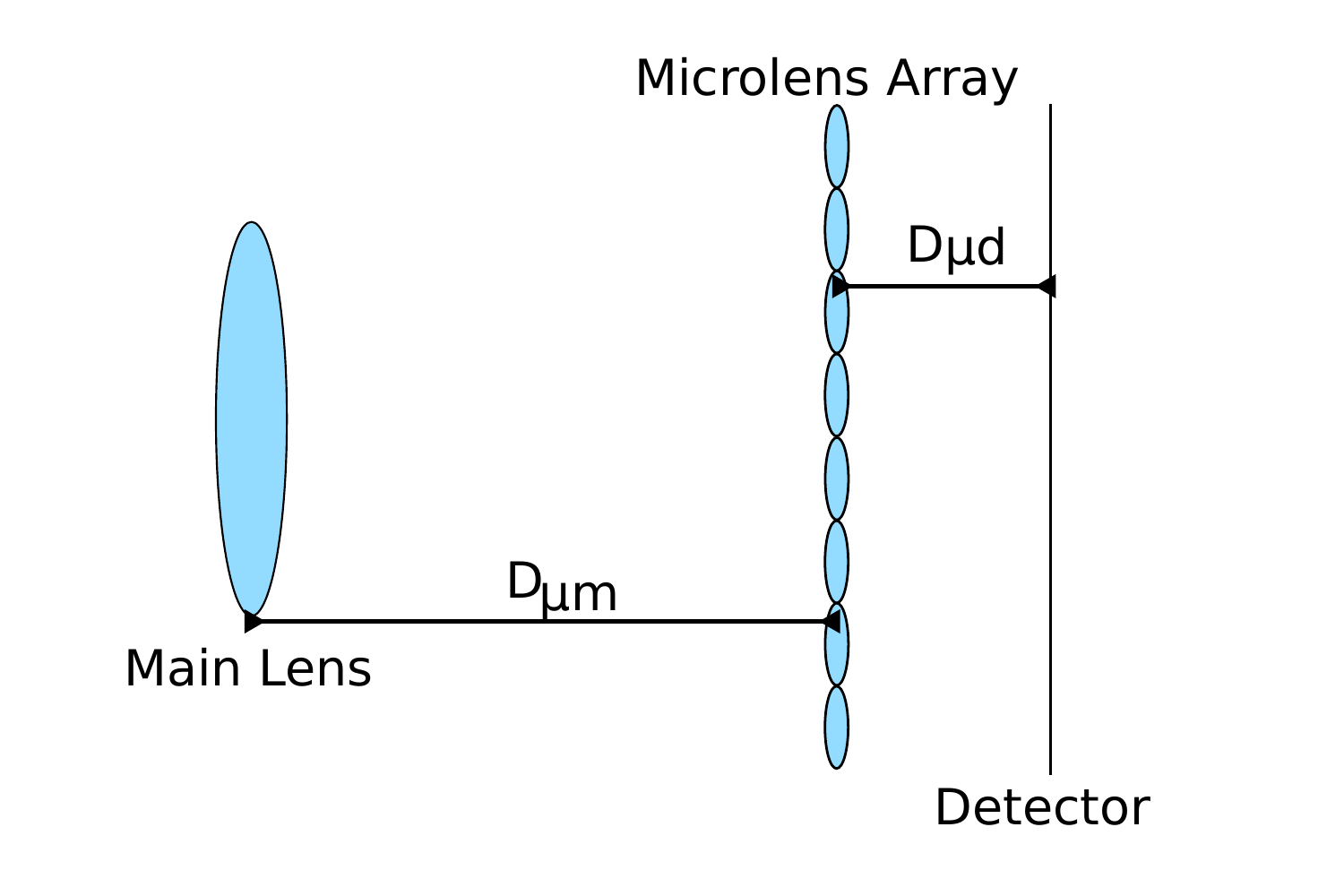}
    \caption{Schematic of a plenoptic camera}
    \label{fig,plenoptic}
\end{figure}

\subsubsection{Plenoptic camera}

In a plenoptic camera, light from the main lens falls on an array of microlenses
that further refract the light onto the detector; see Figure~\ref{fig,plenoptic}.
We model this configuration using two stages:
an initial transport onto the microlens array, followed by masking and transport
through each of the microlenses onto the detector.

Let $D_{\mu \text{m}}$ be the distance between the microlens array and the main
lens.  The optical transformation from the microlens array (prior to refraction
through the microlenses) is the propagation $\bX^{0a} = \bT_{D_{\mu m}}$; let
$\bB^{a s}$ denote the transport operation from the scene $\LL^{s}$ to
the microlens array.  Not all the light incident on the microlens array will be
refracted onto the detector; we assume that light falling between microlenses
is occluded.  We model this with a diagonal masking matrix
$\bM$~\eref{eqn,occlusion}.

The light field incident on the detector, $\LL^d$ is the superposition of the
light that is refracted through each of the $N_\mu$ microlenses;
\ie $\LL^d\GP{\cdot} = \sum_{\mu=1}^{N_\mu} \LL^\mu\GP{\cdot}$.   The optical
transformation from behind each microlens to the angular plane is unique,
because each microlens has a different center location (and possibly different
optical parameters): $\bX^{0\mu} = \X^{0a} \circ \bR_\mu \circ \bT_{D_{d\mu}}$.

Combining the first transport onto the microlens array, masking, and microlens
transport yields the following (noiseless) plenoptic camera measurement model:
\begin{align}
    \y_\text{plenoptic}
    &=
    \sum_{k=1}^K \GB{\sum_{\mu = 1}^{N_\mu} \sqrt{V^\mu} \bB^{d\mu }_k} \M \bB^{\mu s}_k \f^s_k.
\end{align}

\section{Chemiluminescence object model}
\label{sec,chemi}

One application of the computational transport model described above is
reconstructing a 3D chemiluminescence distribution
from images captured by several cameras.  We model the continuous
monochrome
chemiluminscence distribution $f$ using a 3D array of voxels; \ie products of
pillbox functions in the $x$, $y$ and $z$ directions:
\begin{align}
    f\GP{x, y, z}
    &=
    \sum_{j=1}^N
    b\GP{\frac{x - \widetilde x_j}{\Delta_{\mathrm x}}}
    b\GP{\frac{y - \widetilde y_j}{\Delta_{\mathrm y}}}
    b\GP{\frac{z - \widetilde z_j}{\Delta_{\mathrm z}}}
    x_j,
    \label{eqn,voxel}
\end{align}
where $\GC{x_j}$ are the coefficients of the expansion and $\GC{\GP{\widetilde
x_j, \widetilde y_j, \widetilde z_j}}$ are the voxel centers.  To simplify the
acquisition model for a single camera, we assume initially that the object's $x$, $y$ and
$z$ directions are parallel to the camera's $s$, $t$ and axial directions,
respectively.  Section~\ref{sec,rot} describes the generalization for rotated
objects.

We assume that the only source of light in the scene is the chemiluminescent
body $f$, and there are no reflecting objects in the camera's field of view.
Consequently, the radiance along each ray through the scene depends on a
line integral through $f$.

With the object and camera coordinate systems aligned, it is convenient to view
the object $f$ as a set of $N_{\mathrm z}$ ``slices'', each corresponding to a different
value of the axial coordinate $z$.  We ``collapse'' each slice into a light
field at the center of each slice using a small-angle approximation, and use
the light transport tools in the previous section to model the acquisition
process.  Let $\LL^s$ be the light field originating from the slice $z = z_s$,
defined with respect to the camera's angular plane, with vector expansion
coefficients $\w^s_k = \Delta_{\mathrm z} \x^s$, and $\bC^{ds}$ be the camera measurement
operator from $\LL^s$ onto the camera's detector (as described in the previous
section).  The image acquisition model is
\begin{align}
	\y = \sum_{s=1}^{N_{\mathrm z}} \bC^{ds} \w^s.
\end{align}

For some camera models, it is computationally
efficient to factor common terms of the camera operators
$\GC{\bC^{ds}}_{s=1}^{N_z}$.  For example, we factor the plenoptic camera
measurement model into two steps: propagation of each light field view from
every slice onto the microlens array, followed by propagation through the
microlens array:
\begin{align}
	\y_\text{plenoptic}
	&=
	\sum_{k=1}^K
	\underbrace{
		\GB{
			\sum_{\mu=1}^{N_\mu}
			\sqrt{V^\mu}
			\bB^{d\mu}_k
		}
		\M
	}_\text{array onto detector}
	\underbrace{
		\GB{
			\sum_{s=1}^{N_{\mathrm z}}
			\bB^{\mu s}_k
			\w^{s}_k
		}
	}_\text{scene onto array}.
    \label{eqn,plenoptic,factor}
\end{align}
An unfactorized implementation of the plenoptic camera model would require
$O\GP{N_{\mathrm z} N_\mu K}$ transport operations; this
factorization~\eref{eqn,plenoptic,factor} reduces that to $O\GP{K\GP{N_{\mathrm z} +
N_\mu}}$.

\subsection{Rotated perspectives}
\label{sec,rot}

For most multiple-camera or multiple-perspective acquisitions, it is unlikely
that the coordinate systems of each camera and the single object $f$ will
be aligned.  Consequently, we need to be able to image $f$ from a rotated
perspective as well as from the simpler ``head-on'' perspective in the previous
section.  Our approach resamples $f$ and its ``natural'' coefficients $\x$
into a rotated coordinate system with coefficients $\p^\mathrm{r}$ for the
rotated perspective.  The imaging model for each rotated camera is the
composition of the camera model and the rotation operators.

We take an approach similar to the classical three-pass technique used in image
rotation~\cite{unser:95:cbi,paeth:86:afa}.  Similar to those works, we consider
rotations along each axis of less than $45$ degrees; larger rotations can be
modeled as a composition of a permutation operation (\ie rotation by $n$ $90$
degree rotations) and a smaller rotation.

Let $\p \in \Reals^3$ be a point in the object's ``natural'' coordinate
system and $\p^\mathrm{r}$ be the same point in the rotated coordinate system:
$\p = \bTheta \p^\mathrm{r}$, where $\bTheta$ is a $3 \times 3$ rotation matrix.
This matrix can be decomposed as
a composition of a diagonal matrix, $\bD_\bTheta$ and three shear coordinate
transformations:
\begin{align}
	\bTheta
	&=
	\bD_\bTheta
    \bS_{\mathrm z}
    \bS_{\mathrm x}
    \bS_{\mathrm y},
    \label{eqn,rot,decomp}
\end{align}
where $\bD_\bTheta = \diag\GC{D_\mathrm{x},~D_\mathrm{y},~D_\mathrm{z}}$ and \eg
\begin{align}
	\bS_z
	&=
	\GB{
	\begin{matrix}
		1 & 0 & 0 \\
		0 & 1 & 0 \\
        S_\mathrm{zx} & S_\mathrm{zy} & 1
	\end{matrix}
	}.
\end{align}
We implement 3D rotation by applying each of these coordinate
transforms serially, \ie from left to right in~\eref{eqn,rot,decomp}.
Each of the operations is motivated by functional approximation,
as in~\cite{unser:95:cbi} and in~\eref{eqn,xport}.  The first operation,
scaling the coordinates with the diagonal matrix $\bD_\bTheta$,
simply changes the voxel sizes:
\begin{align}
    \Delta_x^r = \frac{\Delta_\mathrm{x}}{D_\mathrm{x}}, \quad
    \Delta_y^r = \frac{\Delta_\mathrm{y}}{D_\mathrm{y}}, \quad
    \Delta_z^r = \frac{\Delta_\mathrm{z}}{D_\mathrm{z}}.
\end{align}
Unlike rotation methods that merely involve
interpolation~\cite{unser:95:cbi,paeth:86:afa}, we we want to maintain the
basis representation~\eref{eqn,voxel} even after rotation.  Thus, to determine the
coefficients $\r^z$ after, say, the $z$ shear transform from the original
coefficients $\x$, we perform a least-squares projection similar to that
in~\eref{eqn,xport} as follows:
\begin{align}
    \r^z
    &=
    \argmin_\r \GN{f\GP{\bS_{\mathrm z}\GP{\cdot}; \r} - f\GP{\cdot; \x}}
    \nonumber \\
    &=
    \bE^{\mathrm z} \x,
\end{align}
where the norm is from $L_2\GP{\Reals^3}$ and
$f\GP{x,y,z; \x}$ is parameterized with the new voxel
sizes~\eref{eqn,voxel}.  In a similar way, we compute $\r^x$ from $\r^z$ and so
on: $\x^r = \bE^{\mathrm y} \bE^{\mathrm x} \bE^{\mathrm z} \x$.

\subsubsection{Shear transformations}

The shear transformations $\bE^{\mathrm x}$, $\bE^{\mathrm y}$ and
$\bE^{\mathrm z}$ are block-diagonal with Toeplitz blocks.  That is, in each
transformation, for all the non-sheared directions (\eg for $\bE^\mathrm{x}$,
for the range of entries with the same $\GP{y, z}$ coordinates), the
transformation is Toeplitz.  The operations also shear only in one direction;
\ie the operators have block structure, \eg
\begin{align}
    \GB{\bE^{\mathrm z}}_{ij}
    &=
    \begin{cases}
        \int_{z \in Z_j} g^{\mathrm z}\GP{z - z_i - \alpha^{x_i y_i}}
            \text{~d}z, & x_i = x_j, y_i = y_j \\
        0, &\text{else.}
    \end{cases}
    \label{eqn,rot,toeplitz}
\end{align}
The interpolation function $g^{\mathrm z}$ is a piecewise quadratic function
and is derived using the same techniques as the light transport expressions
in Section~\ref{sec,model}.  Again, we evaluate the elements of $\bE^{\mathrm z}$
on the fly rather than precomputing and storing them as sparse matrices to
accelerate computation on modern many-core hardware.

\subsection{System models}

All the operations in this section are linear so their composition is also
linear.  For the sake of brevity, we use $\A: \Reals^N \to \Reals^M$ to
denote the composition of operators that represents the action of a
camera that produces an $M$-pixel image from an $N$-voxel chemiluminescence
distribution.

\section{Practical implementation}
\label{sec,practical}

The previous two sections describe the system model that relates the unknown
chemiluminescence expansion coefficients $\x$ to the camera measurements $\y$.
These expressions could be used to precompute the measurement matrix for a
camera, $\A$, that, with the algorithm in Section~\ref{sec,recon}, could be
used to estimate $\x$ from the measurements $\y$.  This is the most common
approach in the
literature~\cite{shroff:13:ifa,schwiegerling:14:pci,bishop:12:tlf,cai:13:pao},
and although it would produce the same results as our proposed method, the
reconstruction would be extremely time-consuming, even with GPU-accelerated
sparse linear algebra libraries.

In X-ray CT~\cite{long:10:3fa} and other time-sensitive inverse problems, this
approach is impractical.  Instead, the entries of $\A$ are
computed on-the-fly by routines that compute the projection
($\A \x$) and backprojection ($\A\tr \y$) matrix-vector products.
In this section, we describe how we implement these operations efficiently
using a GPU for chemiluminescence imaging.

GPUs can execute effectively thousands of ``threads'' simultaneously.  A common
programming model (\eg used by both CUDA and OpenCL) defines an N-dimensional
integer lattice and launches a ``thread'' for each point in the lattice.  The
threads are separated into groups that execute in SIMD (single instruction
multiple data) if possible; \ie the threads execute simultaneously provided the
threads are executing the same instruction (\eg ``add'' or ``multiply'') albeit
on different data.  If two threads in the same group execute different
instructions, the group of threads serially executes each different execution
path; this is called ``thread divergence.'' To maximize throughput, thread
divergence should be avoided as much as possible.

Memory accesses have a similar property.  A group of threads can read or write
simultaneously from the GPU's memory provided the addresses they access are
adjacent (although more modern GPUs relax this requirement); these are called
``coalesced'' memory accesses.  More irregular memory access patterns result in
serialized memory accesses, reducing parallelism.

Finally, to avoid data races or speed-decreasing locking mechanisms, we
assign at most one thread to each coordinate of an output vector.  That is,
when computing the light transport $\f^{p}_k = \bB^{pq}_k \f^{q}_k$, only one
thread writes to each element of $\f^{p}_k$.

\subsection{Adjoint operations}

Sections~\ref{sec,model} and~\ref{sec,chemi} derive expressions for rows of
matrices representing the light transport and rotation operations.  This is
convenient for implementing the forward operations, \eg $\bB^{pq}_k \f^q_k$,
because the elements of $\bB^{pq}_k$ needed by the $i$th thread to compute
$\GB{\bB^{pq}_k\f^q_k}_i = \GB{\GP{\bB^{pq}}_k\tr}_i \tr \f^q_k$ are readily
available.  However, it may be less immediately obvious how to compute the
adjoints needed for the gradient-based optimization method we use in
Section~\ref{sec,recon}.  In some applications, \eg X-ray CT, some system
models may have a parsimonious representation for either its rows or columns
but not both~\cite{long:10:3fa}.  Fortunately, the $L_2$ approximation
framework~\eref{eqn,xport} used here leads to equally efficient
adjoints.  For example, the light transport operation~\eref{eqn,xport} has
adjoint
\begin{align}
    \frac{1}{V^p}\GP{\bB^{pq}}\tr \f^p
    =
    \frac{1}{V^p}\bB^{qp} \f^p
    =
    \frac{V^q}{V^p} \underbrace{\GP{ \frac{1}{V_q} \bB^{qp} \f^p }}_\text{Transport from $p$ to $q$}
    ,
\end{align}
due to the $p, q$ symmetry of the inner product definition of
$\bB^{pq}$~\eref{eqn,xport,ip}; \ie $\bB^{pq} = \GP{\bB^{qp}}\tr$. All the
non-diagonal operations in this paper share this property, so to implement a
fast adjoint we only need a scaled version of a fast forward implementation.

\subsection{Nondiverging coalesced filtering}

The core operation in our models for both light transport and volume rotation
is separable Toeplitz-like~\eref{eqn,xport,sep} or
Toeplitz~\eref{eqn,rot,toeplitz} filtering.  Using separability, we
decompose these operations into two (or three, for rotation) operations.
Although this strategy requires additional kernels to be launched, which incurs
some overhead, the composition-based implementation can reduce computation
significantly.  For example, the separable implementation requires
$O\GP{T + S}$ operations instead of $O\GP{ST}$ operations per pixel, where
$S$ and $T$ are the widths in pixels of the $s$ and $t$ blurs,
respectively.

We store the light field view at the plane $p$ as a $N_{\mathrm s}^p \times N_{\mathrm t}^p$
column-major matrix, \ie the $s$ dimension varies most quickly
in memory.

\begin{table}
    \centering
    \caption{Kernels for light transport}
    \label{tab,kernels}
    \begin{tabular}{llll}
        \toprule
            {\bf Name}
                & {\bf Operation}
                & {\bf Input order}
                & {\bf Output order} \\
        \midrule
            Minor filtering
                & Filter in $t$ direction
                & $N_{\mathrm s}^q \times N_{\mathrm t}^q$
                & $N_{\mathrm t}^p \times N_{\mathrm s}^q$ \\
            Major filtering
                & Filter in $s$ direction
                & $N_{\mathrm t}^p \times N_{\mathrm s}^q$
                & $N_{\mathrm s}^p \times N_{\mathrm t}^p$ \\
        \bottomrule
    \end{tabular}
\end{table}

Table~\ref{tab,kernels} lists the kernels that perform the light
transport operation $\bB^{pq}_k$.  We first filter in the $t$ (minor) direction.
We do this because filtering in the data's major direction would lead to either
thread divergence or noncoalesced memory accesses, since computing integrals
over disjoint regions of the blur
functions~(\ref{eqn,xport,ip},~\ref{eqn,rot,toeplitz}), often involves diverging
operations.  After filtering, each group of threads transposes the data using
local memory before writing it to a temporary vector.  The transpose places $s$
in the minor direction, and we repeat the operation, again ending with a
transpose using local memory to return the data to the original ordering.

\section{Reconstruction algorithm}
\label{sec,recon}

Suppose we have $N_{\mathrm c}$ monochrome cameras with
corresponding system models $\GC{\A_c}$ that acquire
images of the chemiluminescence distribution $\GC{\y_c}$, where
\begin{align}
    \y_c &= \A_c \x + {\bm \varepsilon}_c,
\end{align}
where ${\bm \varepsilon}_c$ denotes the measurement noise for the $c$th
camera.  We assume that the
geometrical configuration of the cameras with respect to the object is known
(which often requires calibration).
Our goal is to estimate $\widehat \x \in \Reals^N$, the vector of basis
expansion coefficients for the chemiluminescence distribution.

Because we may be acquiring data from different types of cameras with
different ADCs and other properties, we do not assume that we know the
relative gains of each of the cameras and will need to estimate them.
Since the rate of photon emission from each voxel via chemiluminescence
is nonnegative, we constrain
$\widehat \x$ to be nonnegative.
This leads to the following penalized nonnegative least squares
problem:
\begin{align}
    \Psi\GP{\x, \GC{\gamma_c}}
    &=
    \frac{1}{2}\GN{\A_1\x - \y_1}_{\W_1}^2
    +
    \sum_{c=2}^{N_{\mathrm c}} \frac{1}{2}\GN{\A_c \x - \gamma_c \y_c}_{\W_c}^2
    \nonumber \\
    &\hspace{8em}+
    \nu \GN{\x}_1
    +
    \mathsf{R}\GP{\x}
    \nonumber \\
    \widehat \x
    &=
    \argmin_{\x \ge \zeros}
    \min_{\GC{\gamma_c}}
    \Psi\GP{\x, \GC{\gamma_c}},
    \label{eqn,pls}
\end{align}
where $\gamma_c$ is the unknown gain for the $c$th camera.  To avoid
degeneracy, we assume the gain for the first camera is unity, without loss of
generality.  The nonnegative diagonal weights matrix $\W_c$ is used to, \eg
select a Bayer pattern of pixels or disregard a damaged region of the detector.
In low-light situations, $\W_c$ could also be used to approximate the combination
of the Poisson photon statistics with the Gaussian electronic readout noise~\cite{snyder:93:irf,snyder:95:cfr}.
The differentiable and convex edge-preserving regularizer $\mathsf{R}$
encourages piecewise smoothness in $\widehat \x$ by penalizing the differences between
adjacent voxels:
\begin{align}
    \mathsf{R}\GP{\x}
    &=
    \frac{\beta}{2}
    \sum_{j=1}^N
    \sum_{l \in \mathcal{N}_j}
    \psi\GP{x_j - x_l},
    \label{eqn,reg}
\end{align}
where $\mathcal{N}_j$ contains the 26 3D neighbors of the $j$th voxel and $\psi$
is an even, convex, differentiable function with bounded curvature of unity.
The optional sparsity-encouraging $\ell_1$ term $\nu \GN{\x}_1$ causes
the algorithm to favor solutions with fewer nonzero entries.

The optimization problem~\eref{eqn,pls} has a convex objective and convex
domain, and there are many algorithms available to solve it.  We use an
iterative shrinkage and thresholding algorithm (FISTA)~\cite{beck:09:afi}
similar to one that has been very effective in accelerating X-ray CT
reconstruction problems~\cite{kim:15:cos} and is summarized in
Table~\ref{tab,alg}.  Appendix~\ref{sec,hessian} demonstrates the majorization
condition underlying this algorithm.

\begin{table}
    \centering
    \caption{Reconstruction algorithm}
    \label{tab,alg}
    \fbox{%
        \begin{minipage}{3in}
            Initialize: $\x\snz = \zeros$; $\z\snz = \zeros$;
                $\gamma_1 = 1$,
                $\gamma_c = 0$ for $c = 2, \ldots, N_c$,
                $t\snz = 1$.

            Compute the majorizer
            $\D = \diag_j \GC{ \GB{\sum_{c=1}^{N_c} \A_c \tr \W_c \A_c \ones}_j }$.

            Loop for $n = 1,\ldots,N_\text{iter}$:
            \begin{enumerate}
                \item In parallel for $c = 1, \ldots, N_c$:
                    \begin{enumerate}
                        \item Compute the projection $\p_c \leftarrow \A_c \z\sn$
                        \item If $c > 1$, compute the gain
                            $$\gamma_c \leftarrow \frac{\y_c\tr \W_c \p_c}{\y_c \tr \W_c \y_c}.$$
                        \item Compute the gradient for camera $c$:
                            $$\g_c \leftarrow \A_c\tr\W_c\GP{\p_c - \gamma_c \y_c}.$$
                    \end{enumerate}
                \item Accumulate the camera gradients,
                    $\g \leftarrow \sum_{c=1}^{N_c} \g_c$.
                \item Update:
                    \begin{align}
                        t &= \frac{1 + \sqrt{1 + 4\GP{t\sn}^2}}{2}
                        \nonumber \\
                        \w
                        &=
                        \GB{\z\sn - \GP{\D + 26\beta\I}^{-1}
                            \GP{\g + \grad\mathsf{R}\GP{\z\sn}}}_+
                        \nonumber
                        \\
                        x\snp_j &= \GB{w_j - \nu}_+
                        \nonumber \\
                        \z\snp &= \x\snp + \frac{t\sn - 1}{t\snp}\GP{\x\snp - \x\sn}
                    \end{align}
            \end{enumerate}

            Output: $\x^{(N_\text{iter})}$.
        \end{minipage}
    }
\end{table}

The most time consuming steps in the algorithm described by Table~\ref{tab,alg}
are the projection and backprojection steps for each camera.  Fortunately,
these can be computed in either partially in parallel by the out-of-order
execution capability in modern GPUs or completely using multiple GPUs or
multiple computers on a network.  In our experiments with a GPU with 2.5 GB of
memory, we stored all variables in single-precision floating point format on
the GPU, and the algorithm did not require any host-GPU transfers except to
compute the inner products needed for camera gain estimation.

\subsection{View subset acceleration}
\label{sec,subset}

To further accelerate the reconstruction, we use an approximation similar to the
``ordered subsets'' approximation from X-ray CT~\cite{erdogan:99:osa,kim:15:cos} or the
stochastic gradient approach from machine learning.  Instead of computing the
exact gradient of the data fidelity term for each camera, we compute an approximation
using only a subset of the angular plane:
\begin{align}
  \g_c^\text{exact}
  &=
    \sum_{k=1}^K \A_{ck}\tr\W_c\GP{\A_{ck}\x - \gamma_c \y_k}
  \nonumber \\
  &\approx
  \frac{K}{\GM{\mathcal{S}_n}}
  \sum_{k \in \mathcal{S}_n}
  \A_k\tr\W_c\GP{\frac{K}{\GM{\mathcal{S}_n}}\A_{ck}\x - \gamma_c \y_k}
  = \g_c^\text{approx.}
  \label{eqn,subset}
\end{align}
The subset of the angular plane $\mathcal{S}_n \subset \GC{1, \ldots, K}$ is
iteration-dependent and heuristically chosen such
that the approximation~\eref{eqn,subset} is reasonably accurate.

In our experiments below, we divide the angular plane's basis functions into
$N_\text{subset}$ disjoint subsets formed by taking every $N_\text{subset}$th
view from the angular plane ordered lexicographically.  Alternative orders have
been proposed for X-ray CT~\cite{guan:94:apa} and are an area of possible
future research.

\subsection{Algorithm parameters}

There are two types of parameters in the proposed algorithm: regularizer parameters
and parameters related to the camera system models $\GC{\A_c}$.

The edge-preserving regularizer $\mathsf{R}$~\eref{eqn,reg} has two parameters:
the nonnegative weight $\beta$ and the potential function $\psi$.  The
regularizer we use is common in image reconstruction, and there are many
options for choosing these parameters in the literature,
\eg~\cite{ramani:12:rps}.  In our experiments, we used the simple quadratic
potential function $\psi\GP{t} = \frac{1}{2}t^2$ and, due to the early nature
of the experiments we performed, chose the weight $\beta$ by coarse hand
tuning.  Further work could certainly refine these choices.

The light transport chain in the camera system models $\GC{\A_c}$ are defined
relative to optical planes parameterized by the angular basis function $a$ and
the plane's spatial discretization.  The computational complexity of performing
a projection or backprojection operation is linear in the number of points $K$ in
the angular plane's discretization, but more fine discretizations potentially yield more
accurate models for light transport physics.  Similarly, the pillbox angular
basis function requires slightly more computation, but offers a more accurate
model than the Dirac angular basis function.  Finally, the degree of subset-based
acceleration (Section~\ref{sec,subset}) has an effect on accuracy as well.

\section{Experiments and results}
\label{sec,experiment}

This section reports four experiments:
\begin{itemize}
    \item Section~\ref{sec,angular} explores the tradeoff in computation time
        and image quality
        for different discretizations of the angular plane.

    \item Section~\ref{sec,single} shows a simulated reconstruction from a
        single plenoptic camera with a hexagonal microlens configuration.  
        The reconstructed chemiluminescence
        distribution has very poor depth resolution, although we can recover
        some depth information by thresholding the reconstruction.

    \item Section~\ref{sec,phantom} illustrates reconstructing a simulated
        phantom from three cameras with known geometric configuration: a single
        plenoptic camera and two simple cameras at $\pm 30$
        degrees.  The additional axial information provided by the two
        secondary cameras yields a reconstruction with much higher
        fidelity than the single-camera experiment in Section~\ref{sec,single}.

    \item Finally, Section~\ref{sec,flame} reports a flame reconstruction from
        three views taken by a single Raytrix R29 plenoptic camera~\cite{perwass:12:sl3} of a flame
        on a rotating stage, emulating a three-camera acquisition.
\end{itemize}
For all the experiments, we used an implementation of the system and object
models described in previous sections, and the reconstruction algorithm in
Section~\ref{sec,recon}.  All software was implemented in OpenCL and the
Rust programming language and run on a NVIDIA Quadro K5200 GPU with 2.5GB
of memory.  Source
code and configuration files for the first three experiments will be available
under an open source license.

\begin{table}
    \centering
    \caption{Simulated plenoptic camera configuration}
    \label{tab,gorgon}
    \begin{tabular}{ll}
        \toprule
        {\bf Property} & {\bf Value} \\
        \midrule
        Main lens focal length & 105 mm \\
        Main lens radius & 4.5 mm \\
        \midrule
        Distance between main lens and array, $D_{\mu m}$ & 112.0 mm \\
        Distance between array and detector, $D_{\mu d}$ & 2.2 mm \\
        \midrule
        Microlens radius & 100 $\mu$m \\
        Microlens focal lengths & 2.8 mm, 3.0 mm, 3.2 mm \\
        Microlens pattern & Hexagonal \\
        \midrule
        Detector pixel pitch & 5 $\mu$m \\
        Detector dimensions & $2048 \times 2048$ pixels \\
        \bottomrule
    \end{tabular}
\end{table}

\subsection{Model parameters}
\label{sec,angular}

Finer discretizations of a camera's angular plane may yield more accurate models
but require higher computational cost.  At a certain point this reaches a point of
diminishing returns and the higher accuracy is no longer worth
further increases in complexity, particularly given other approximations
such as neglecting aberrations.  This section explores that tradeoff using the
phantom and plenoptic camera used in the reconstruction in
Section~\ref{sec,phantom}; see Table~\ref{tab,gorgon} for the camera parameters
and Figure~\ref{fig,plenoptic} for a cartoon of the camera.

\begin{figure}
    \centering
    \begin{subfigure}[b]{1.6in}
        \centering
        \includegraphics[width=\textwidth]{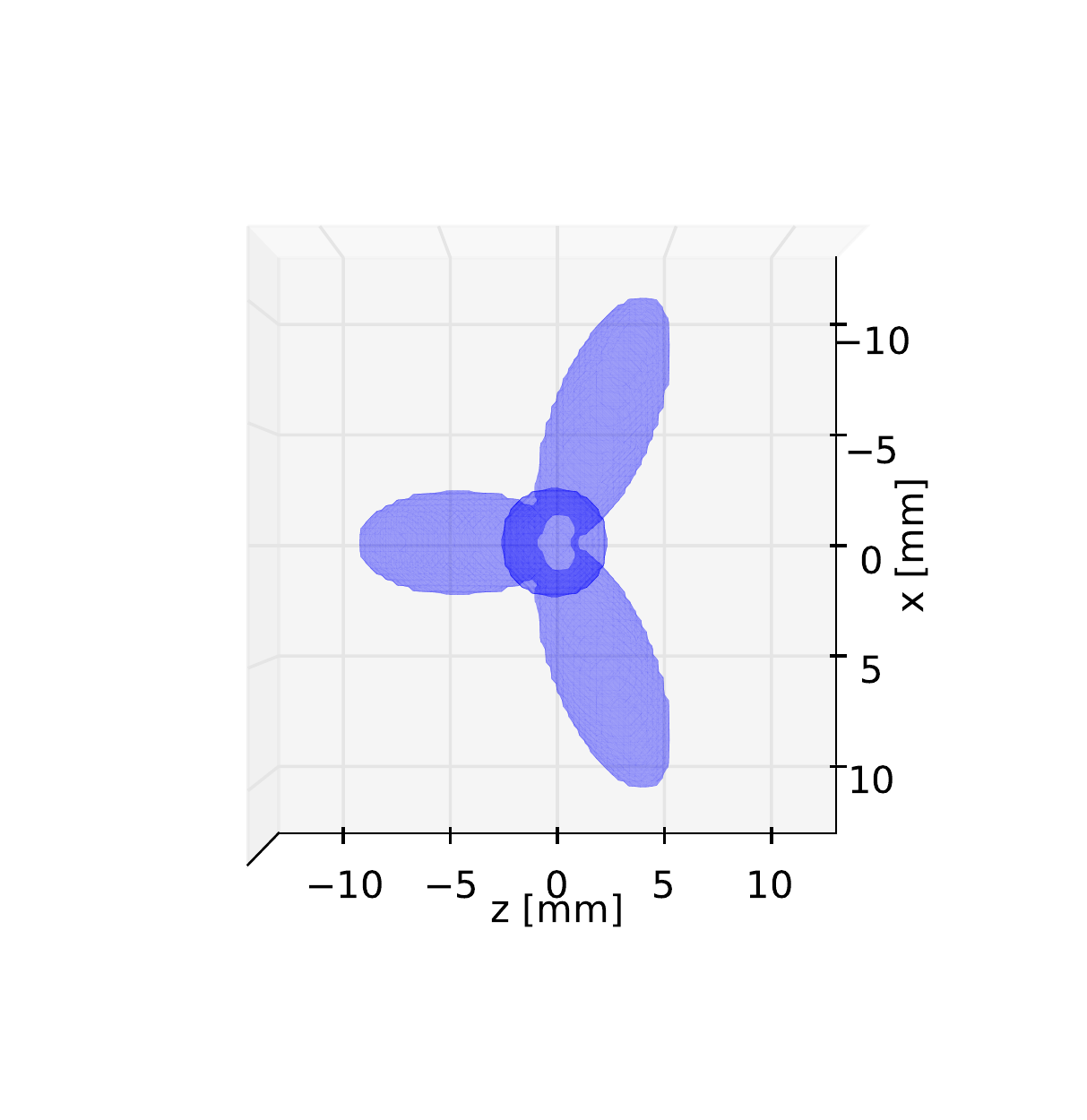}
        \caption{Top}
    \end{subfigure}
    \begin{subfigure}[b]{1.6in}
        \centering
        \includegraphics[width=\textwidth]{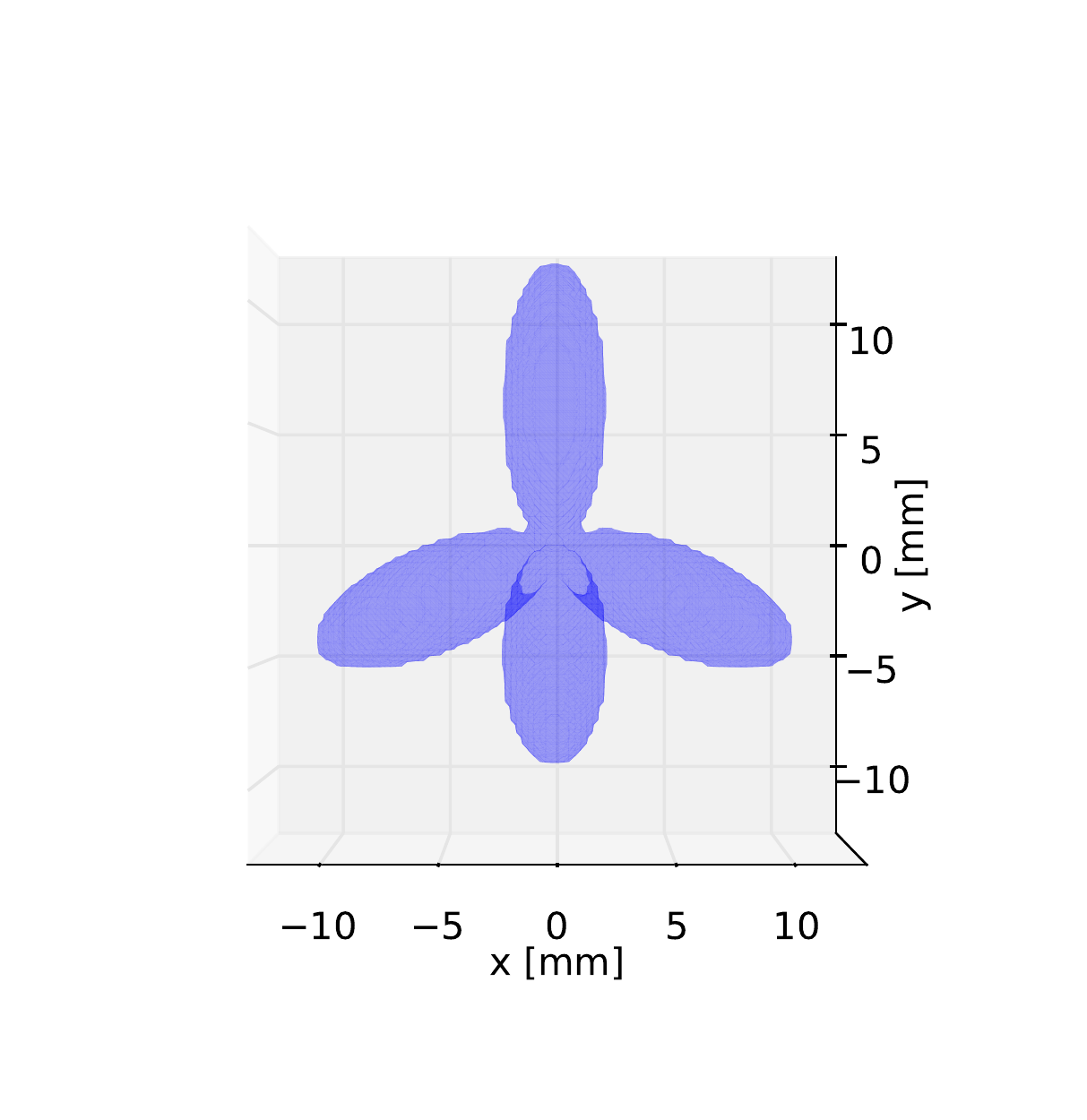}
        \caption{Axial}
    \end{subfigure}

    \begin{subfigure}[b]{1.6in}
        \centering
        \includegraphics[width=\textwidth]{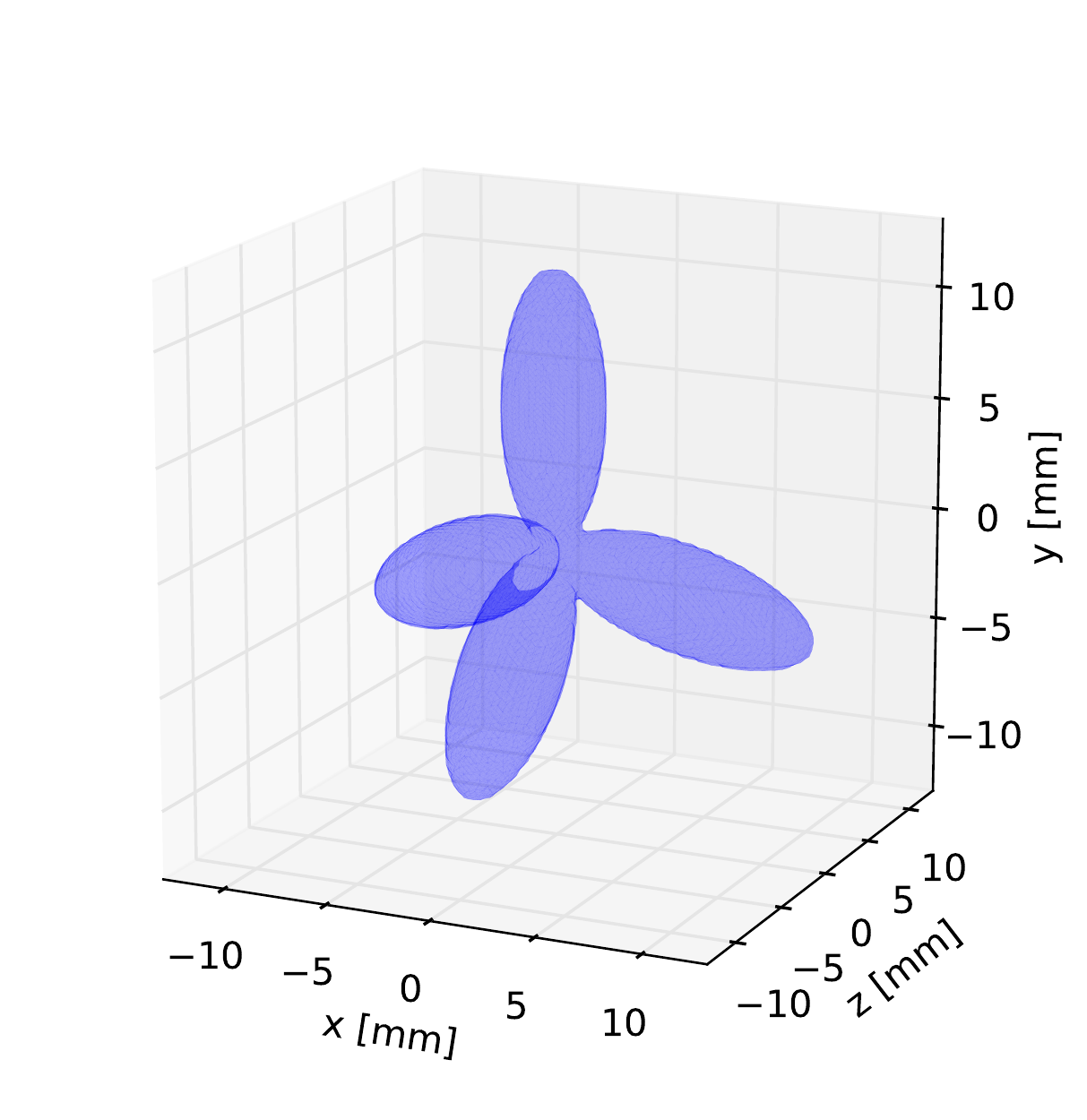}
        \caption{Perspective}
    \end{subfigure}
    \begin{subfigure}[b]{1.6in}
        \centering
        \includegraphics[width=\textwidth]{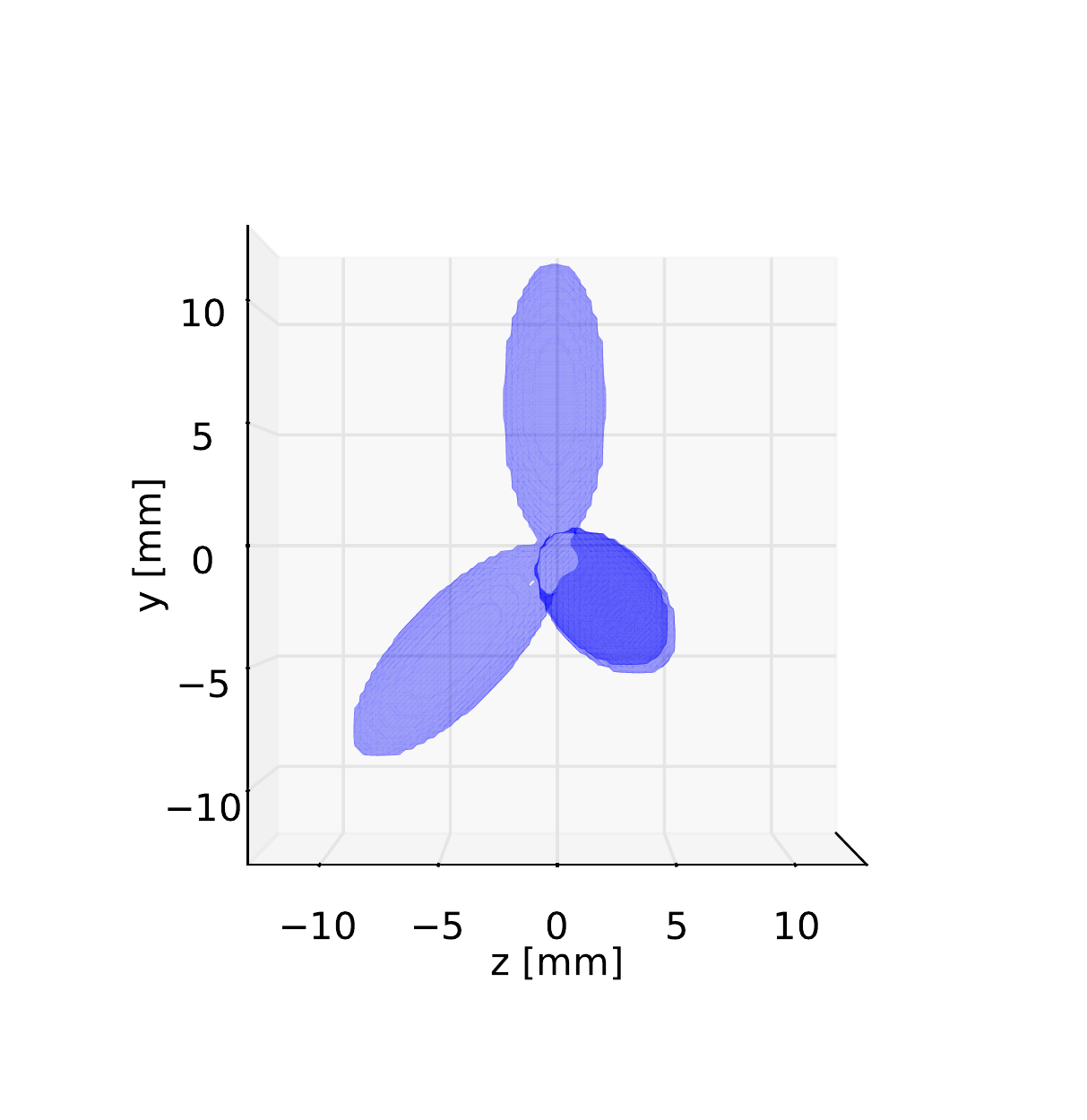}
        \caption{Side}
    \end{subfigure}

    \caption{The four-pronged phantom used in the simulations.}
    \label{fig,phantom,render}
\end{figure}

\begin{figure*}
    \centering
    \includegraphics[width=7in]{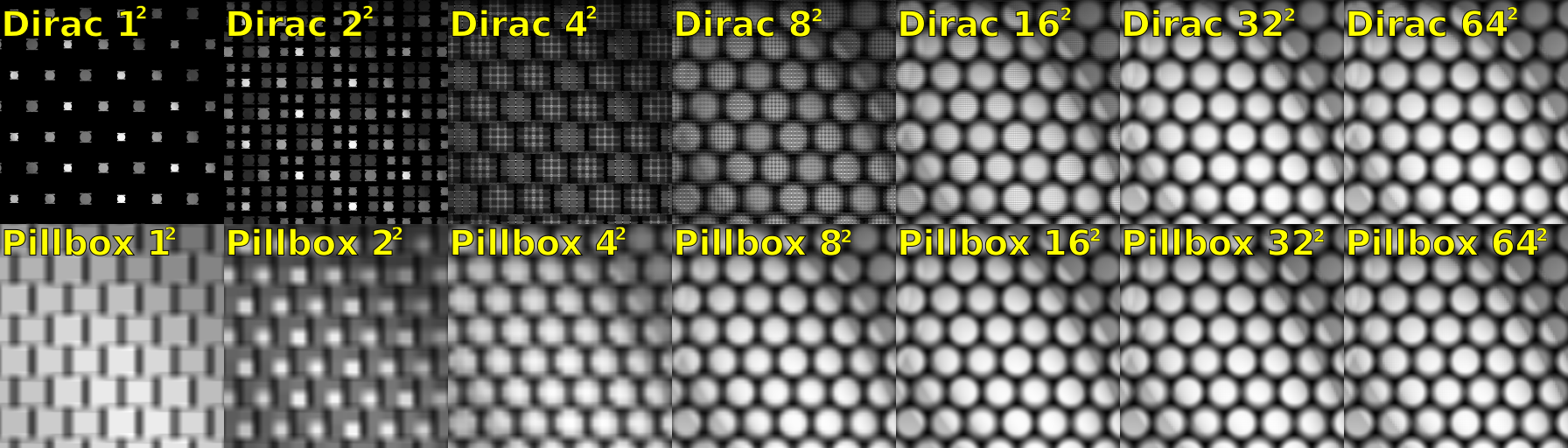}
    \caption{$256 \times 256$-pixel subimages from the simulated plenoptic
        camera with different angular discretizations.  As the angular
        discretization increases, both models produce high quality images,
        but the pillbox basis function is more accurate at coarser
        discretizations.}
    \label{fig,collage}
\end{figure*}

Figure~\ref{fig,phantom,render} shows the four-pronged phantom we imaged
in this experiment, and
Figure~\ref{fig,collage} shows $256 \times 256$ subimages from the
simulated plenoptic camera at each parameterization of the angular
plane we tested.  Each number in the parameterization gives the
coarseness of the discretization along the $s$ and $t$ directions;
\eg ``Pillbox $8^2$'' means a $8 \times 8$ discretization of the
angular plane using the pillbox angular basis function.

Both the Dirac and pillbox models produce high quality images at higher angular
discretizations, but the Dirac model yields significant aliasing artifacts at
lower discretizations.  This is a well-known weakness of ``pinhole camera''
modeling from image-based rendering methods~\cite{lin:00:otn,chai:00:ps}, and
is mitigated by using the pillbox basis function.

Figure~\ref{fig,param,accuracy} plots the normalized difference,
\begin{align*}
    \text{NSD}\GP{\y}
    &=
    \frac{\GN{\y - \y_\text{high-quality}}^2}{\GN{\y_\text{high-quality}}^2}
\end{align*}
of each
configuration to the highest-quality pinhole rendering, and confirms the
superior accuracy of the pillbox angular plane discretization over the pinhole
discretization.  Since the higher quality rendering of the pillbox comes at
essentially no additional computational time over the Dirac approach, as
Figure~\ref{fig,param,time} shows, we use the pillbox-based model for the
following reconstructions.

\begin{figure}
    \centering
    \begin{subfigure}[b]{1.5in}
        \centering
        \includegraphics[width=\textwidth]{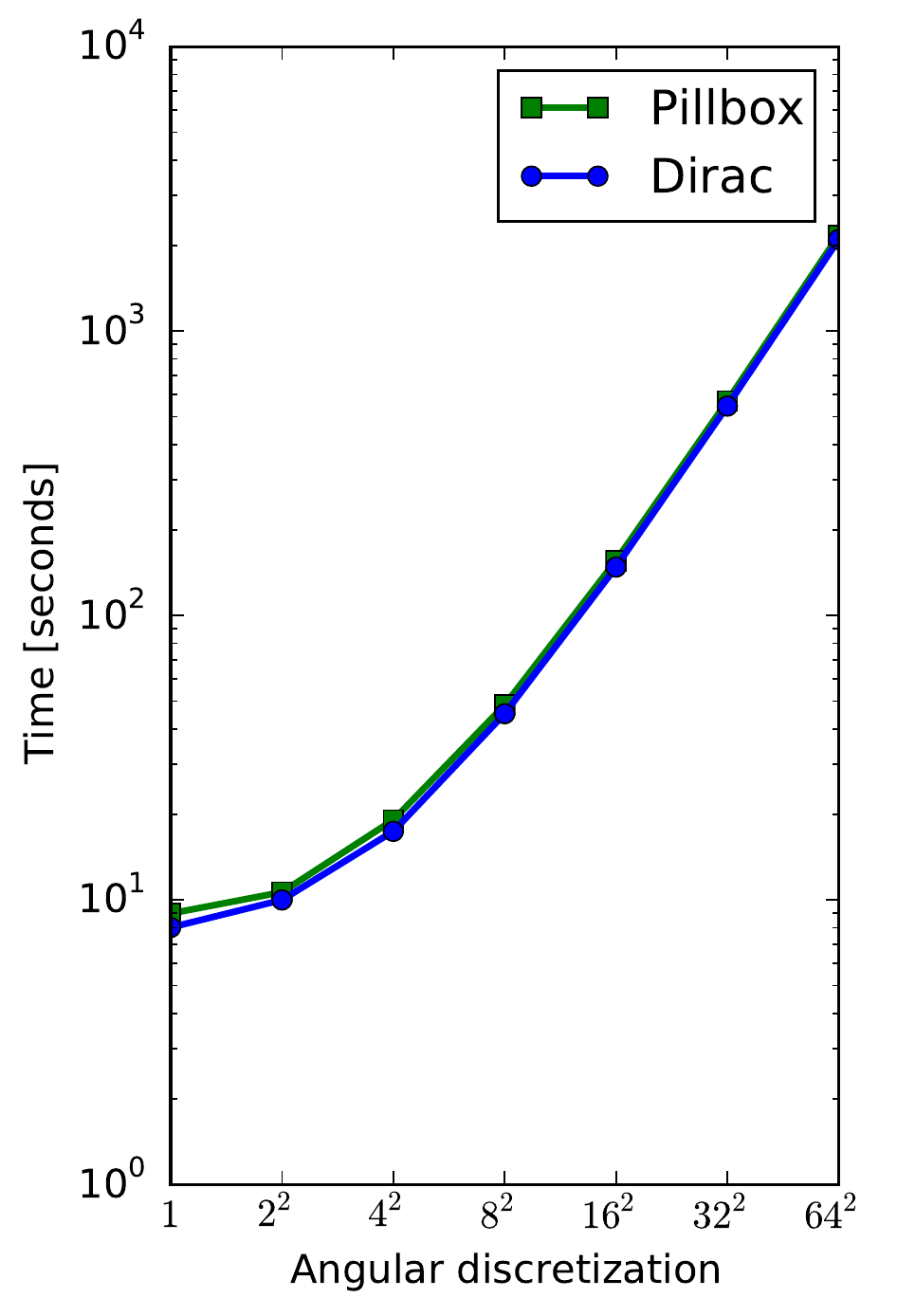}
        \caption{Projection time}
        \label{fig,param,time}
    \end{subfigure}
    \begin{subfigure}[b]{1.5in}
        \centering
        \includegraphics[width=\textwidth]{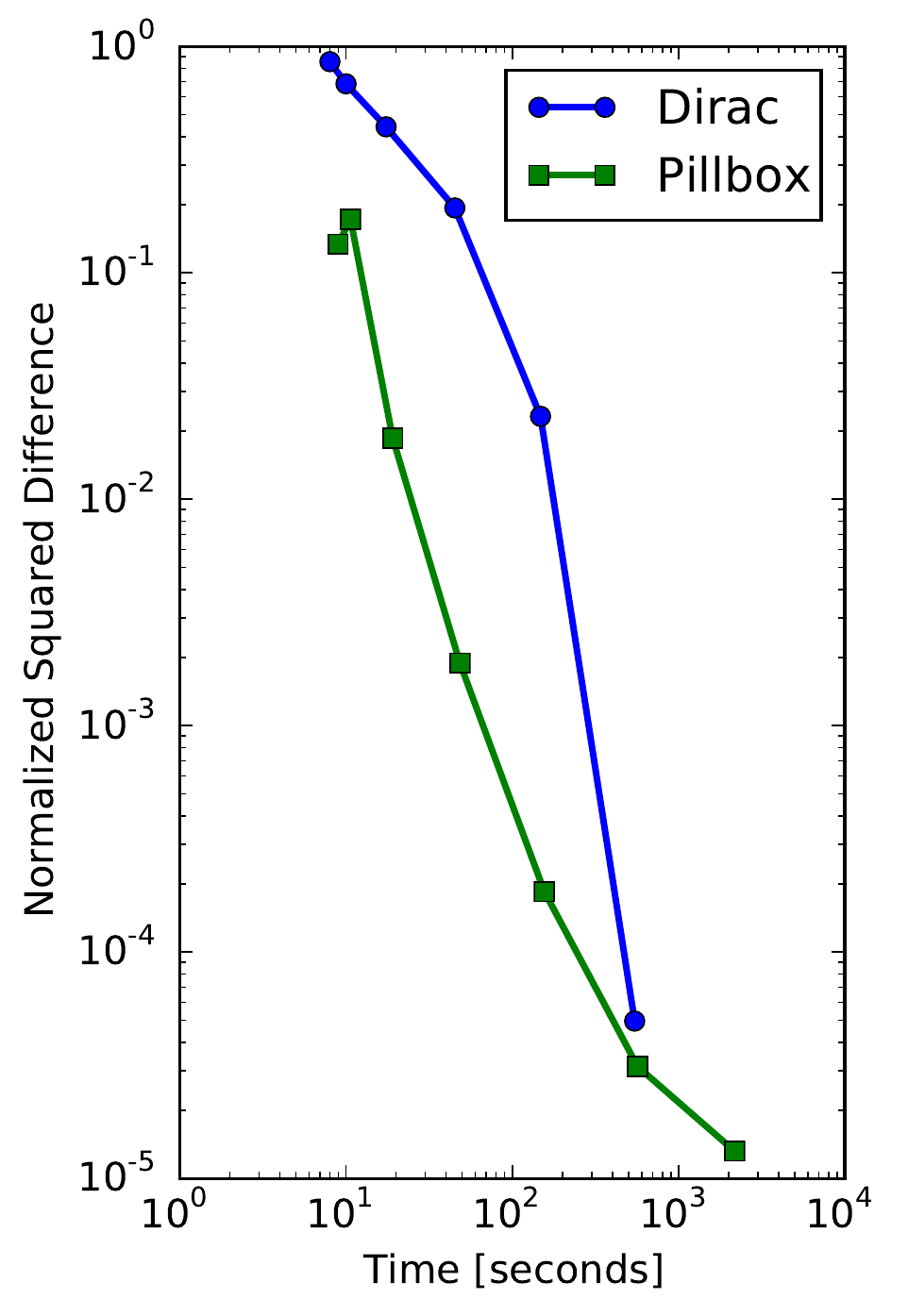}
        \caption{Projection accuracy}
        \label{fig,param,accuracy}
    \end{subfigure}
    \caption{Projection times and NSD for a the simulated plenoptic camera
        model.  NSD is computed relative to the image from the Dirac
        image with $64 \times 64$ angular discretization.  The pillbox
        model is essentially as fast as the Dirac model, but produces
        more accurate results, particularly at coarser discretizations.}
    \label{fig,param,plots}
\end{figure}

\subsection{Single camera reconstruction}
\label{sec,single}

\begin{figure}
    \centering
    \begin{subfigure}[b]{1.6in}
        \centering
        \includegraphics[width=\textwidth]{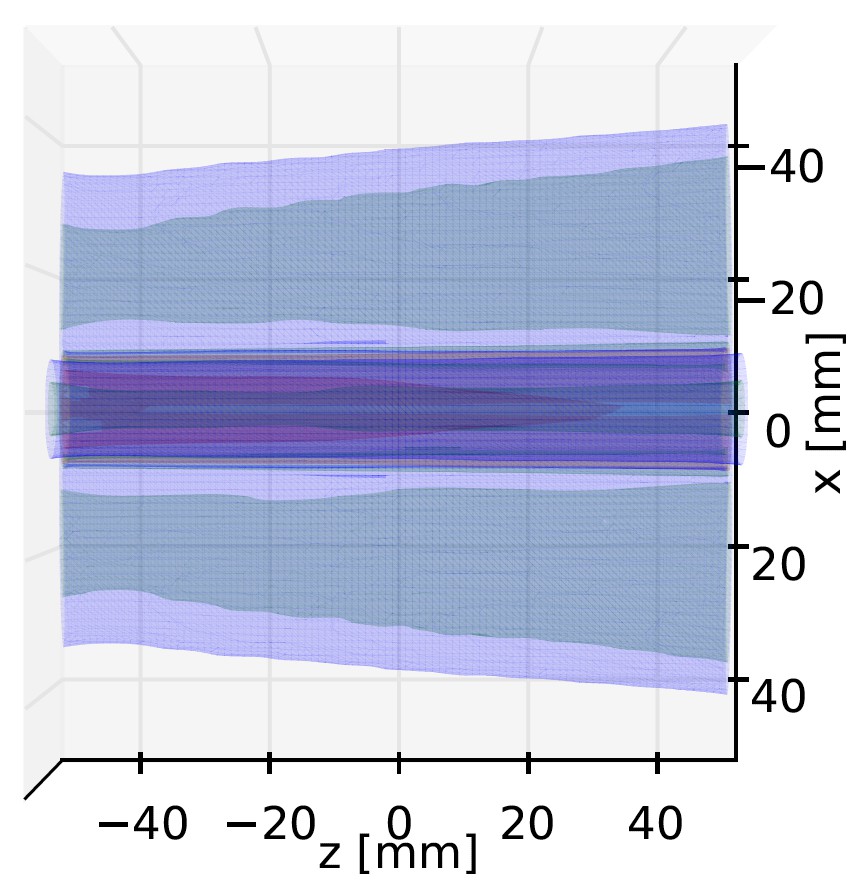}
        \caption{Top}
    \end{subfigure}
    \begin{subfigure}[b]{1.6in}
        \centering
        \includegraphics[width=\textwidth]{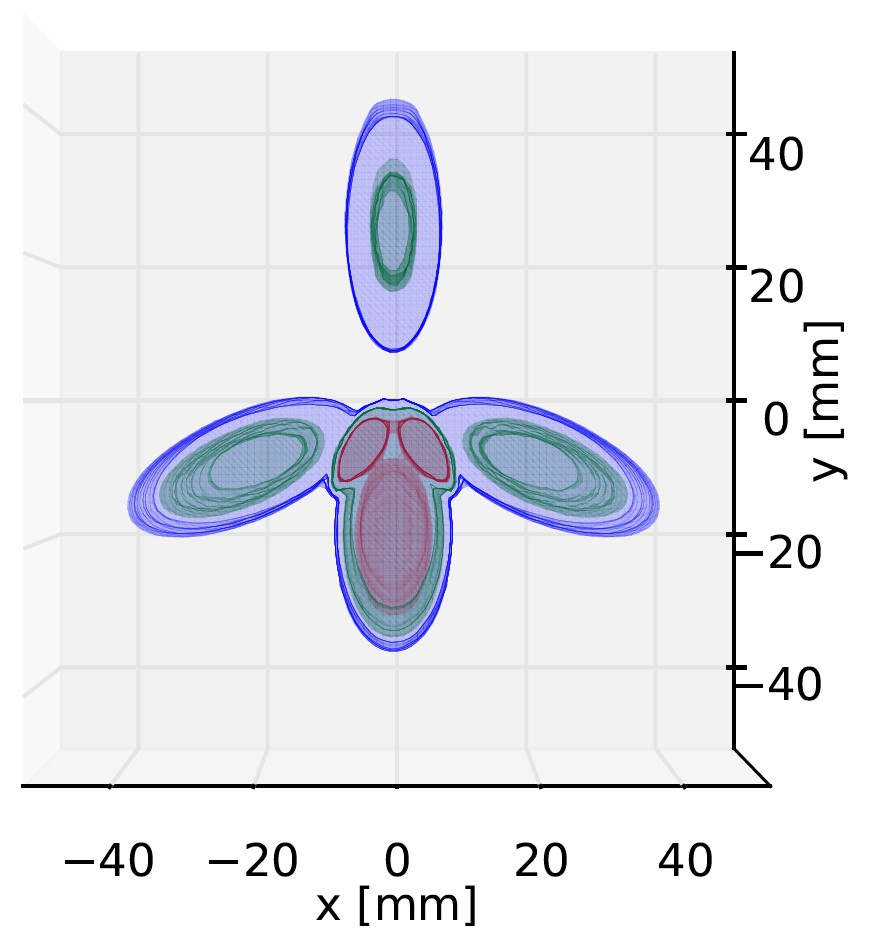}
        \caption{Axial}
    \end{subfigure}

    \begin{subfigure}[b]{1.6in}
        \centering
        \includegraphics[width=\textwidth]{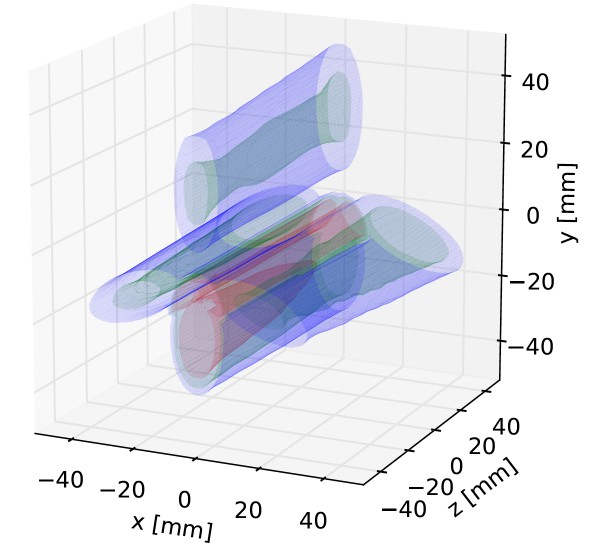}
        \caption{Perspective}
    \end{subfigure}
    \begin{subfigure}[b]{1.6in}
        \centering
        \includegraphics[width=\textwidth]{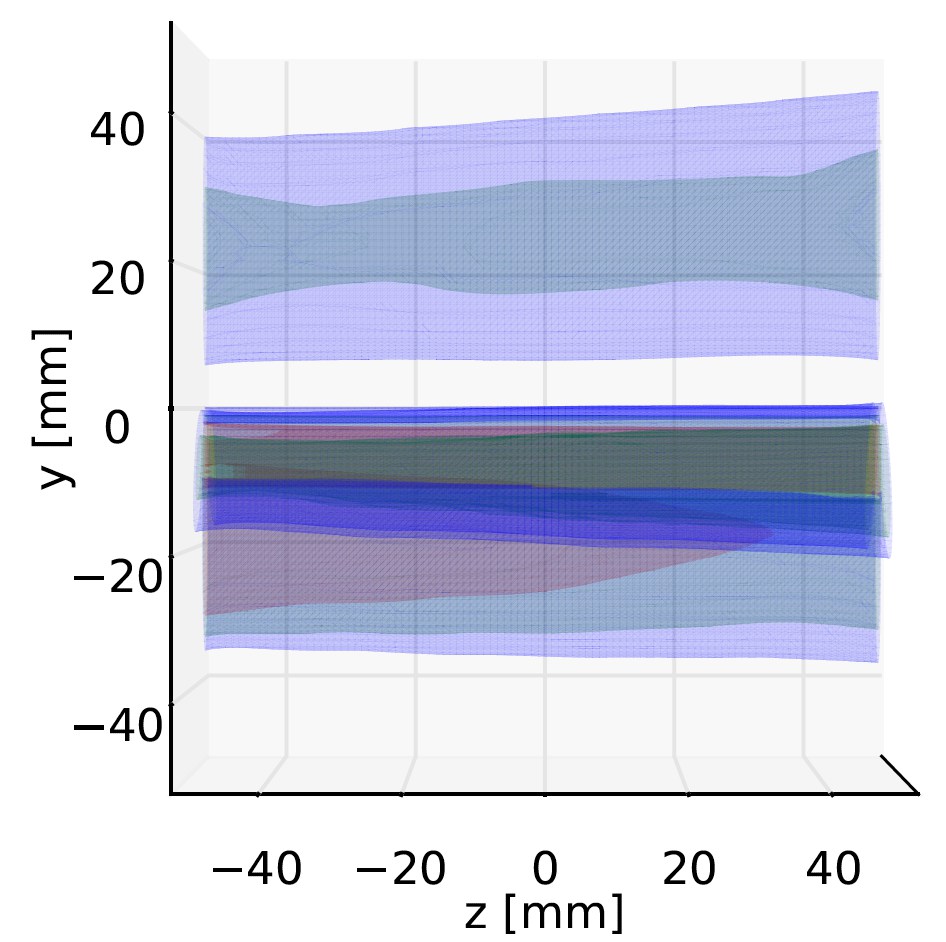}
        \caption{Side}
    \end{subfigure}

    \caption{Simulated reconstruction from one plenoptic camera.
        The figures show level sets at 45\% (blue), 60\% (green) and 75\% (red)
        of the reconstruction's maximum value.  The reconstruction has very
        poor axial ($z$) resolution.
        }
    \label{fig,single,render}
\end{figure}

We simulated image an of the four-pronged phantom taken by the plenoptic 2.0
camera in Table~\ref{tab,gorgon}, shown in Figure~\ref{fig,phantom,gorgon},
using the pinhole basis function and a $64 \times 64$ angular plane
discretization.  We then reconstructed the image using a $16 \times
16$ angular discretization using the pillbox basis function and
$N_\text{subset}=2$ subset acceleration.  The image is reconstructed onto a
$100 \times 100 \times 100$ volume with $1~\text{mm}^3$ voxels; the data were
generated from a $200 \times 200 \times 200$ volume with $\frac{1}{8} \text{mm}^3$
voxels to avoid an ``inverse crime.'' Each image update (subset) took about 70
seconds, and we ran the algorithm for 160 iterations.

Figure~\ref{fig,single,render} shows level sets from the reconstructed image.
The reconstruction has good transaxial ($xy$) resolution, but very poor axial
($z$) resolution.  The reconstruction quality is too poor to be useful; this
confirms earlier findings~\cite{nien:15:mbi} using another system
model~\cite{bishop:12:tlf} and similar reconstruction approach.  We posit that
the poor reconstruction quality is due to very limited angular information;
the next section adds two regular cameras at $\pm 30^\circ$ to augment the
missing angular information.

\subsection{Three camera reconstruction}
\label{sec,phantom}

\begin{figure}
    \centering
    \begin{subfigure}[b]{1.6in}
        \centering
        \includegraphics[width=\textwidth]{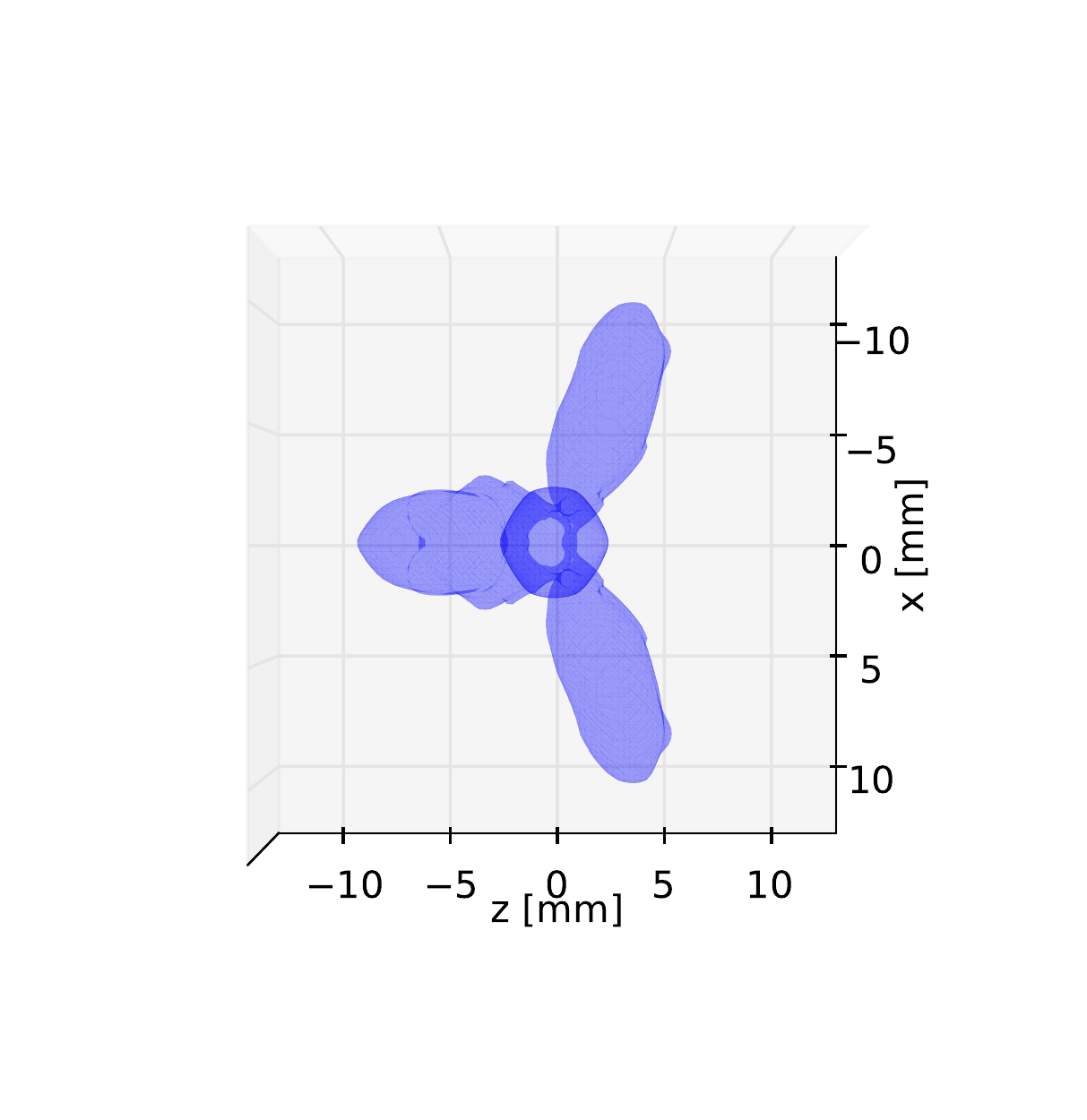}
        \caption{Top}
    \end{subfigure}
    \begin{subfigure}[b]{1.6in}
        \centering
        \includegraphics[width=\textwidth]{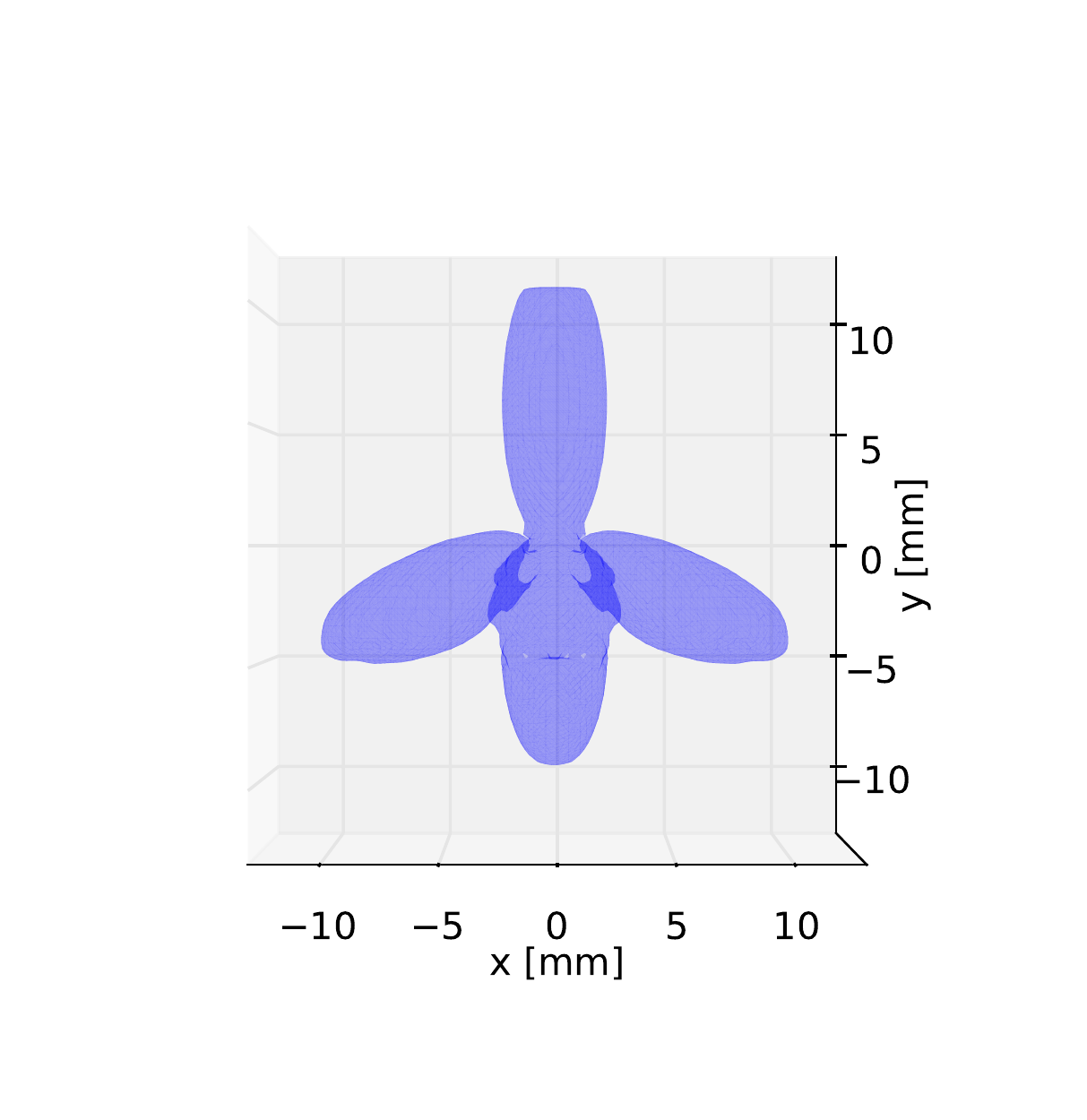}
        \caption{Axial}
    \end{subfigure}

    \begin{subfigure}[b]{1.6in}
        \centering
        \includegraphics[width=\textwidth]{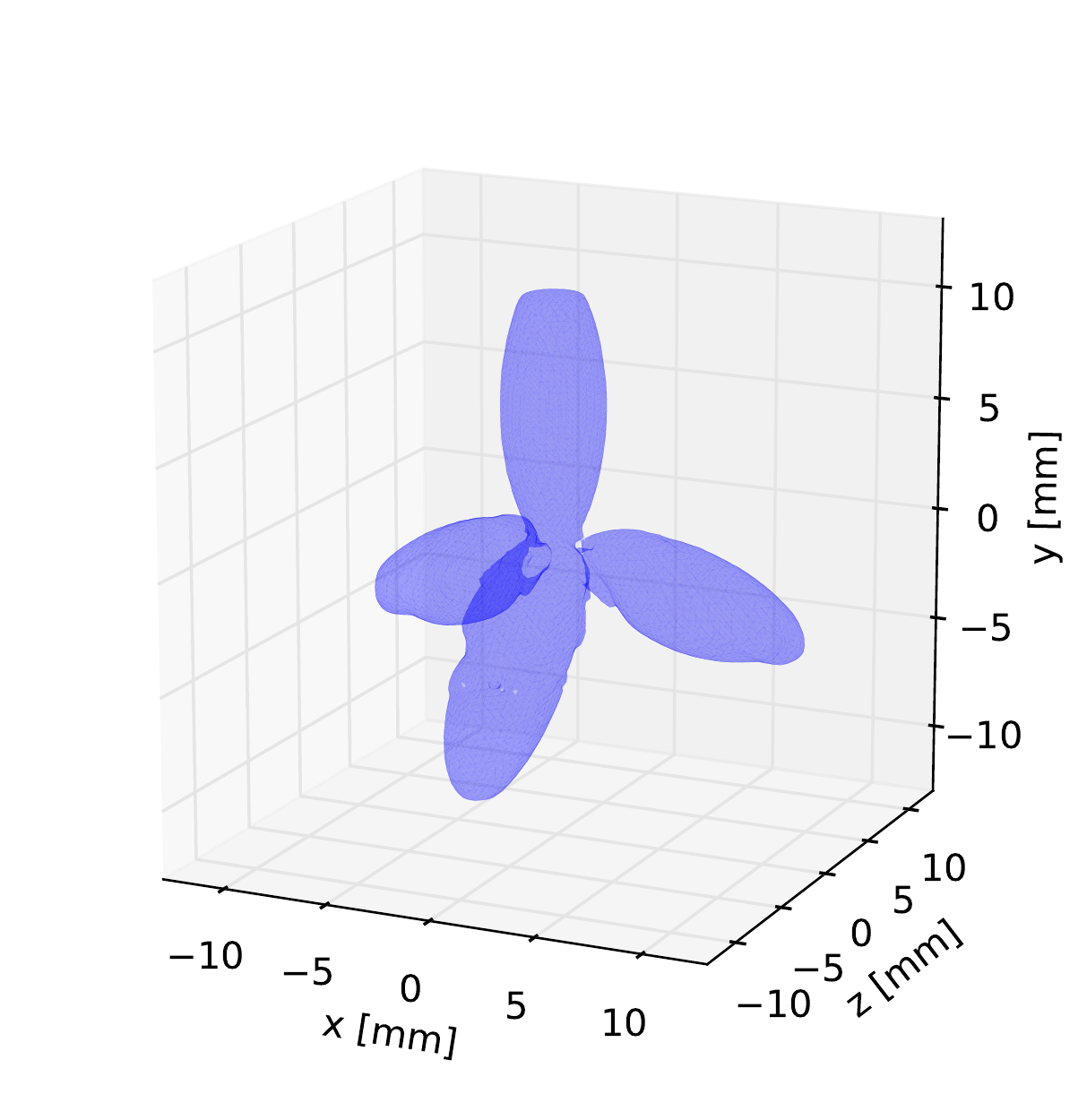}
        \caption{Perspective}
    \end{subfigure}
    \begin{subfigure}[b]{1.6in}
        \centering
        \includegraphics[width=\textwidth]{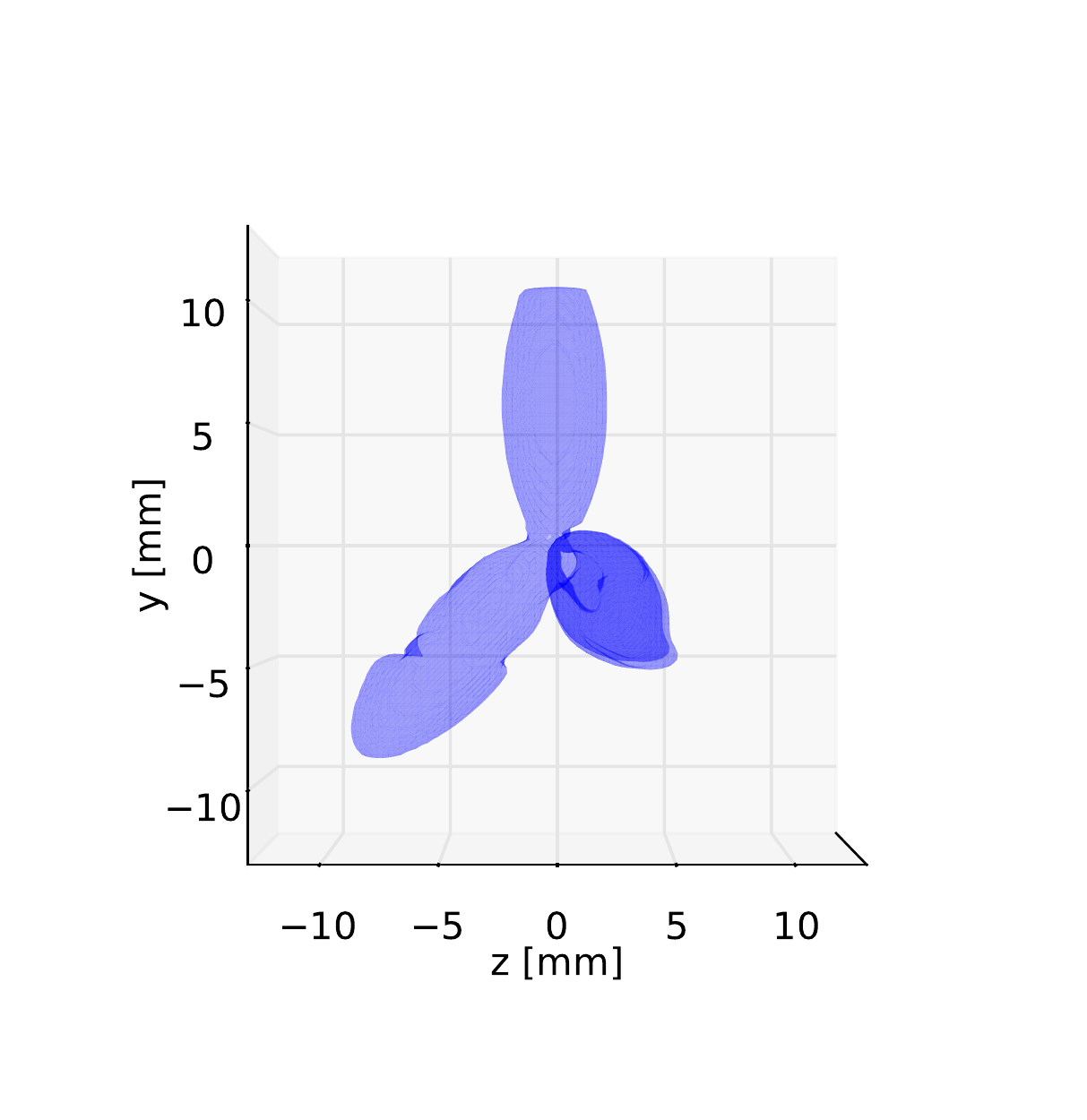}
        \caption{Side}
    \end{subfigure}

    \caption{Simulated reconstruction from one plenoptic camera and two
        cameras.  The figures show level sets at 40\% of the 
        reconstruction's maximum value.  The additional angular information
        from the two secondary cameras results in a high quality 
        reconstruction.}
    \label{fig,g3,render}
\end{figure}

\begin{table}
    \centering
    \caption{Simulated single-lens camera configuration}
    \label{tab,single}
    \begin{tabular}{ll}
        \toprule
        {\bf Property} & {\bf Value} \\
        \midrule
        Lens focal length & 30 mm \\
        Lens radius & 5 mm \\
        \midrule
        Distance between lens and detector, $D$ & 31.3 mm \\
        \midrule
        Detector pixel pitch & 5 $\mu$m \\
        Detector dimensions & $1024 \times 1024$ pixels \\
        \bottomrule
    \end{tabular}
\end{table}

We rendered images of the phantom from $\pm 30^\circ$ using the single lens
camera described in Table~\ref{tab,single}.  Figure~\ref{fig,phantom,views}
shows the three images used to perform the reconstruction.  With the additional
axial information, we found that a coarser angular discretization was acceptable
for reconstruction: we used a $8 \times 8$ pillbox discretization for each
camera's angular plane and no subset acceleration.  Each iteration took about
49 seconds to run, and we ran 100 iterations of the proposed algorithm, but
the iterates were stable after about 50 iterations.

Figure~\ref{fig,g3,render} shows the level sets of the reconstructed phantom.
The additional angular information from the single lens cameras has a dramatic
effect: the reconstructed images are much higher fidelity reconstructions of
the original phantom in Figure~\ref{fig,phantom,render}.  The proposed camera
model also handles multiple cameras, gain estimation, and several camera
types effectively.  The next section, we validates the proposed technique
using real data collected with a plenoptic camera and a rotational stage.

\begin{figure}
    \centering
    \begin{subfigure}[b]{1in}
        \centering
        \includegraphics[width=\textwidth]{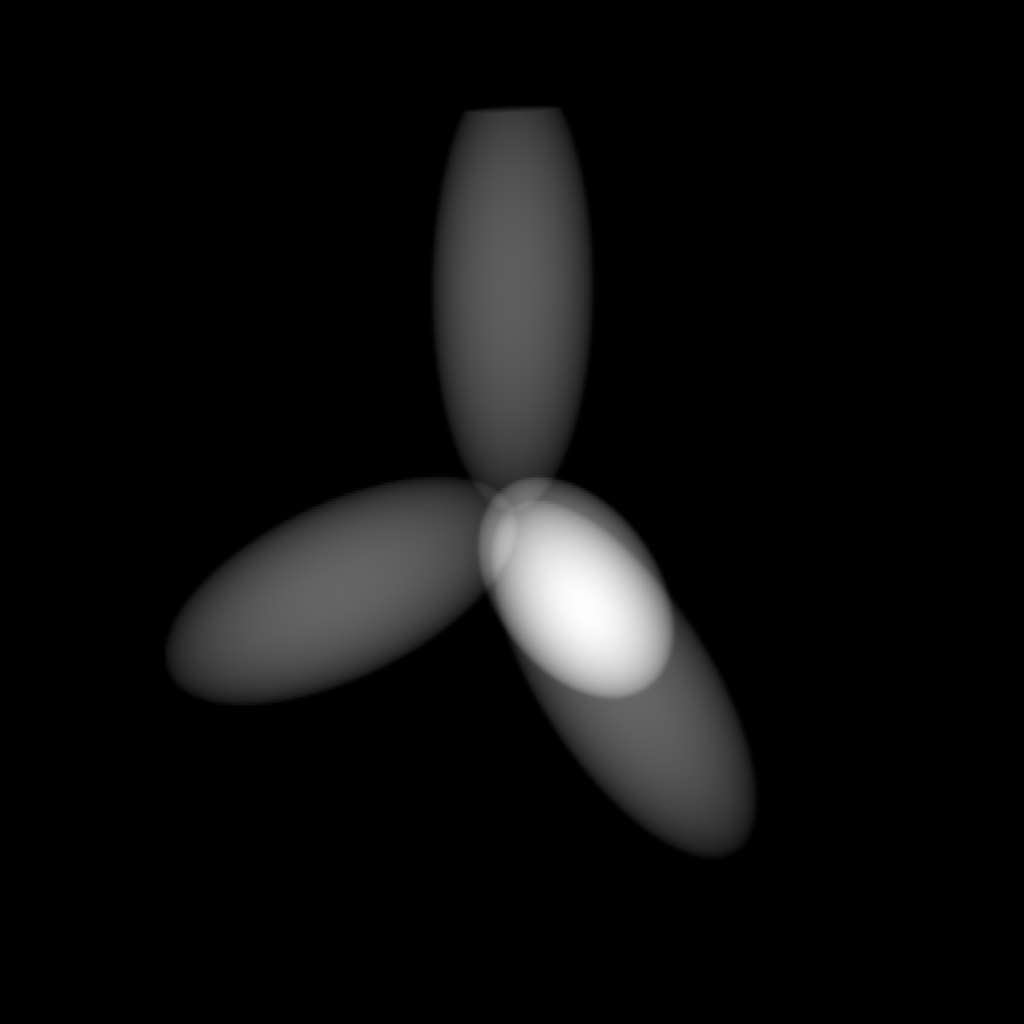}
        \caption{$-30^\circ$ view}
        \label{fig,phantom,left}
    \end{subfigure}
    \begin{subfigure}[b]{1in}
        \centering
        \includegraphics[width=\textwidth]{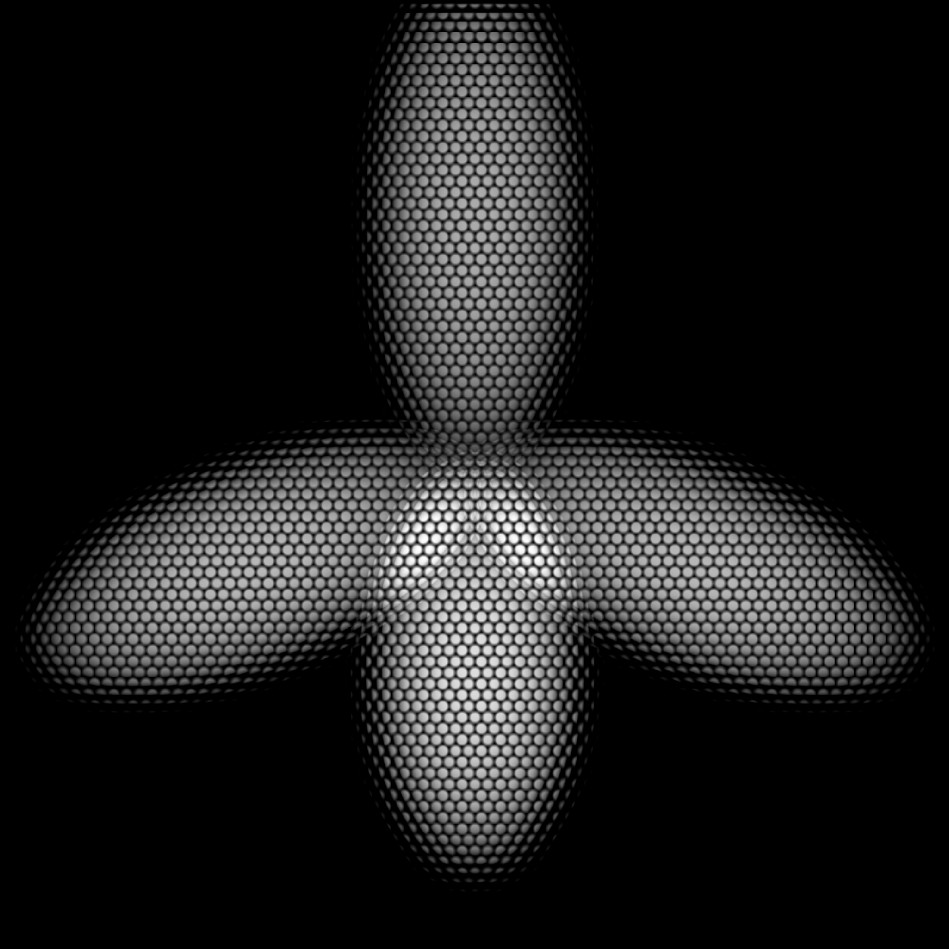}
        \caption{Plenoptic image}
        \label{fig,phantom,gorgon}
    \end{subfigure}
    \begin{subfigure}[b]{1in}
        \centering
        \includegraphics[width=\textwidth]{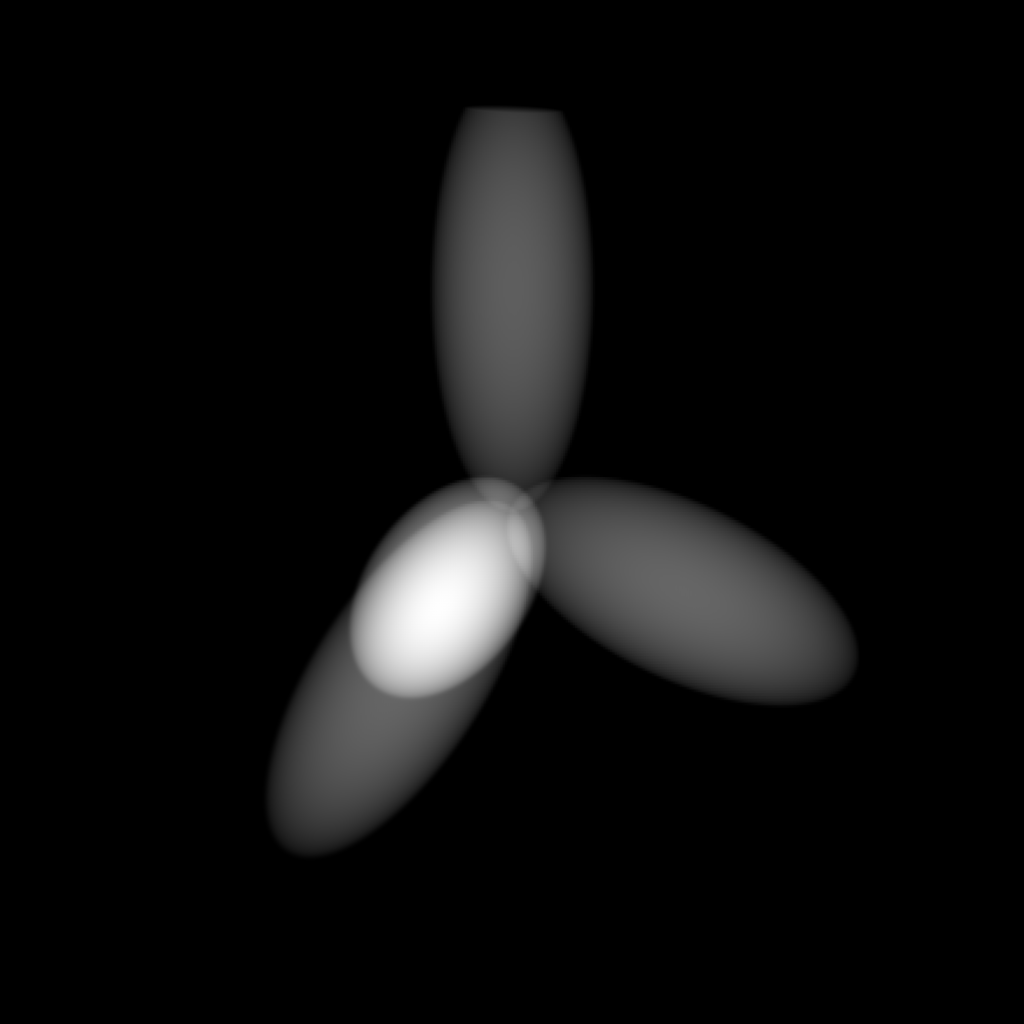}
        \caption{$+30^\circ$ view}
        \label{fig,phantom,right}
    \end{subfigure}
    \caption{Three simulated views of a four-pronged torch phantom.}
    \label{fig,phantom,views}
\end{figure}

\subsection{Flame reconstruction}
\label{sec,flame}

\begin{figure}
    \centering
    \begin{subfigure}[b]{1in}
        \centering
        \includegraphics[width=\textwidth]{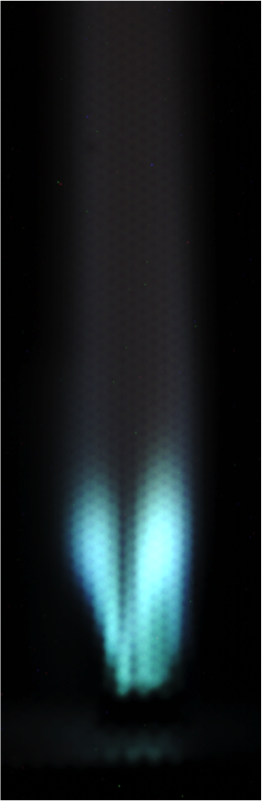}
        \caption{$-15^\circ$}
    \end{subfigure}
    \begin{subfigure}[b]{1in}
        \centering
        \includegraphics[width=\textwidth]{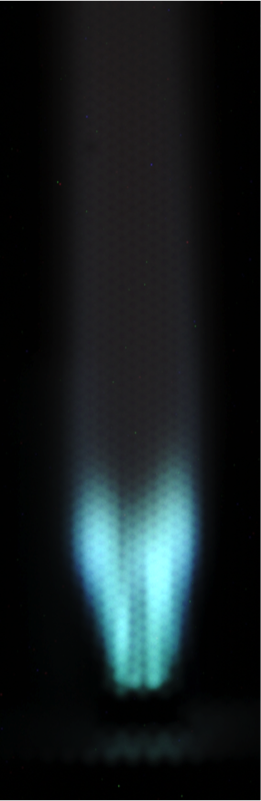}
        \caption{Center}
    \end{subfigure}
    \begin{subfigure}[b]{1in}
        \centering
        \includegraphics[width=\textwidth]{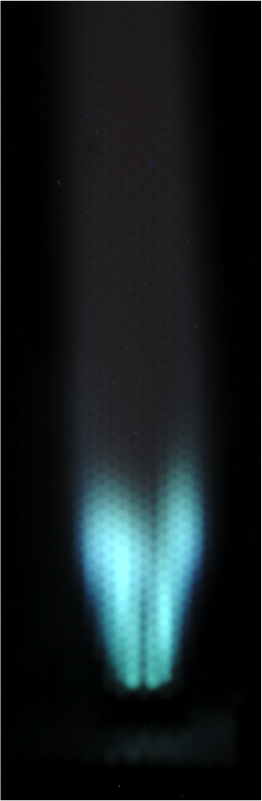}
        \caption{$+15^\circ$}
    \end{subfigure}
    \caption{Subsets of refocused images created using Raytrix software 
        for the flame reconstruction experiment.  These images are for illustration;
        our reconstruction used the raw Bayer-encoded images from the camera.}
    \label{fig,real,views}
\end{figure}

\begin{figure}
    \centering
    \begin{subfigure}[b]{1.6in}
        \centering
        \includegraphics[width=\textwidth]{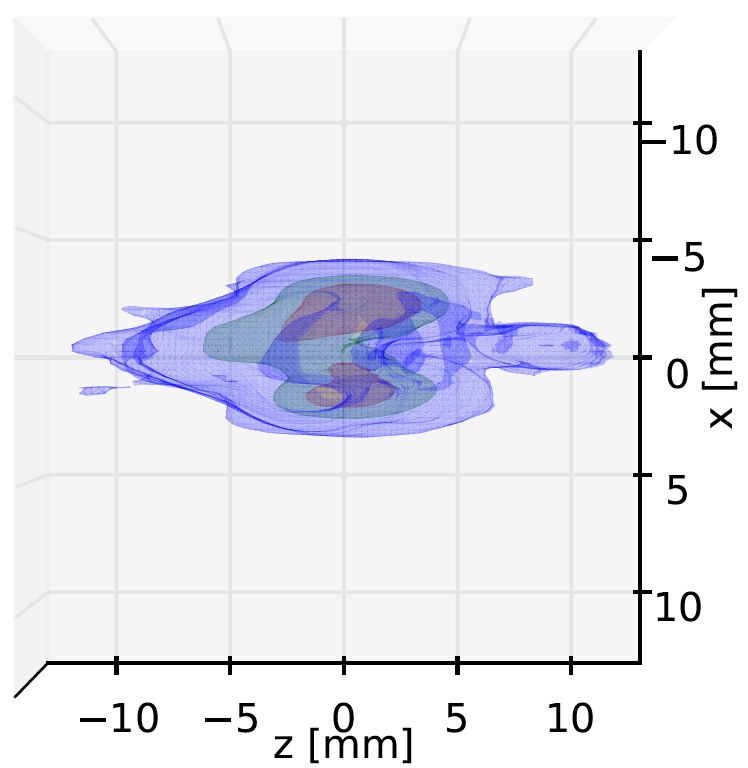}
        \caption{Top}
    \end{subfigure}
    \begin{subfigure}[b]{1.6in}
        \centering
        \includegraphics[width=\textwidth]{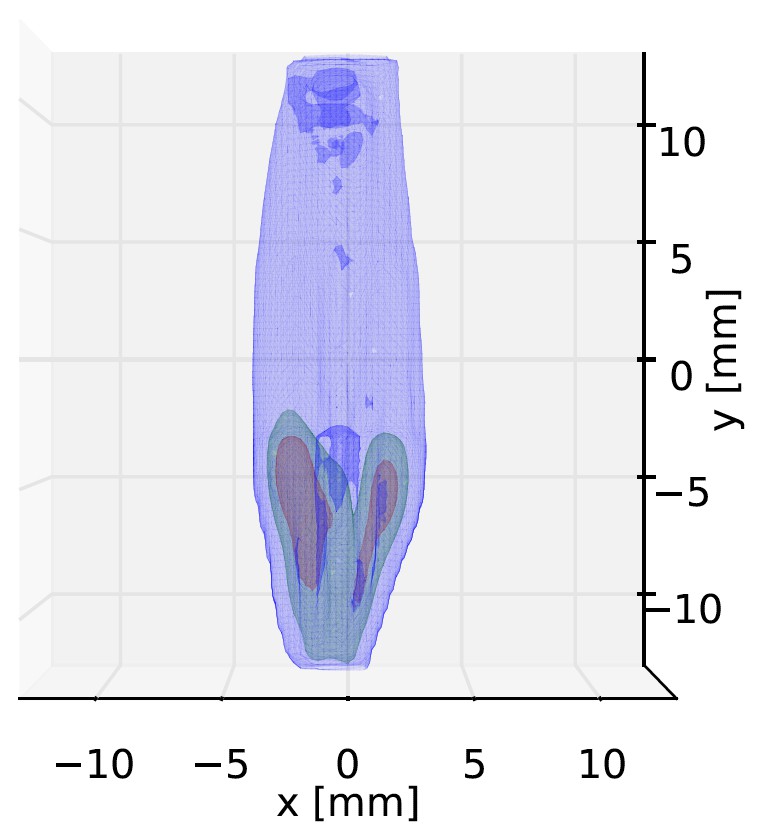}
        \caption{Axial}
    \end{subfigure}

    \begin{subfigure}[b]{1.6in}
        \centering
        \includegraphics[width=\textwidth]{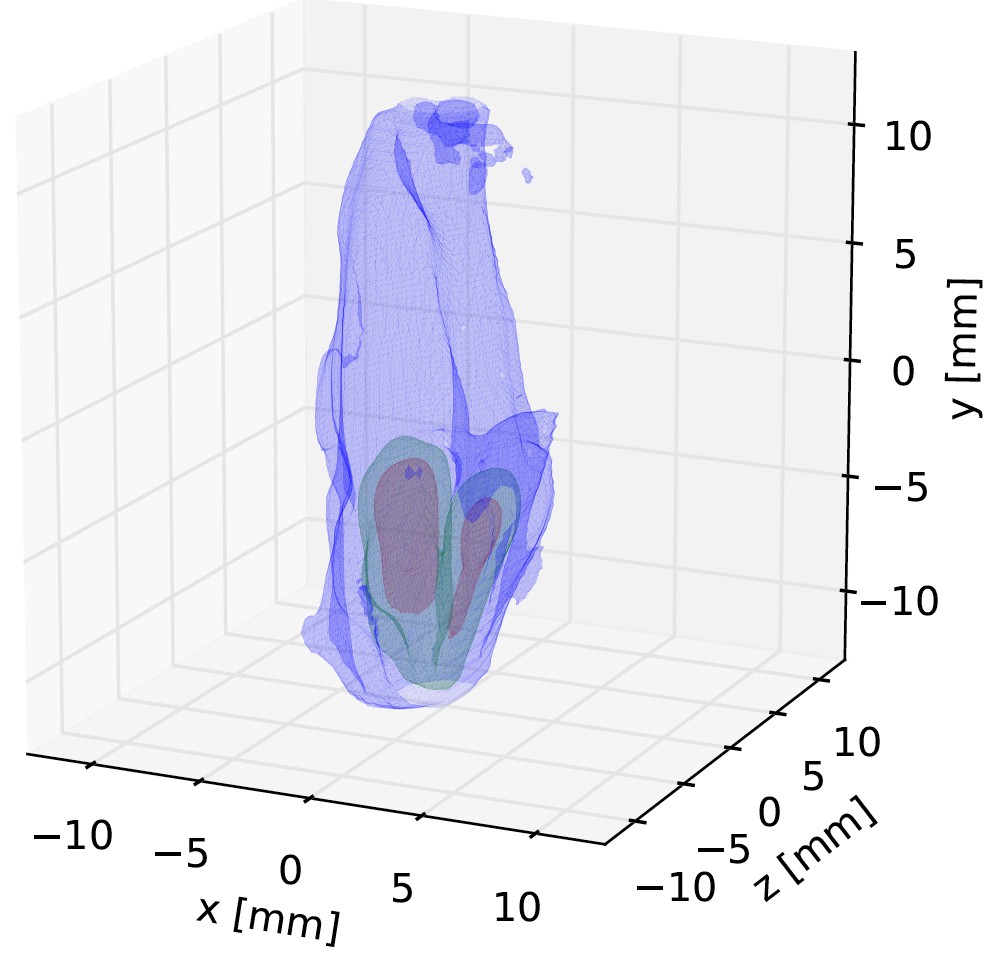}
        \caption{Perspective}
    \end{subfigure}
    \begin{subfigure}[b]{1.6in}
        \centering
        \includegraphics[width=\textwidth]{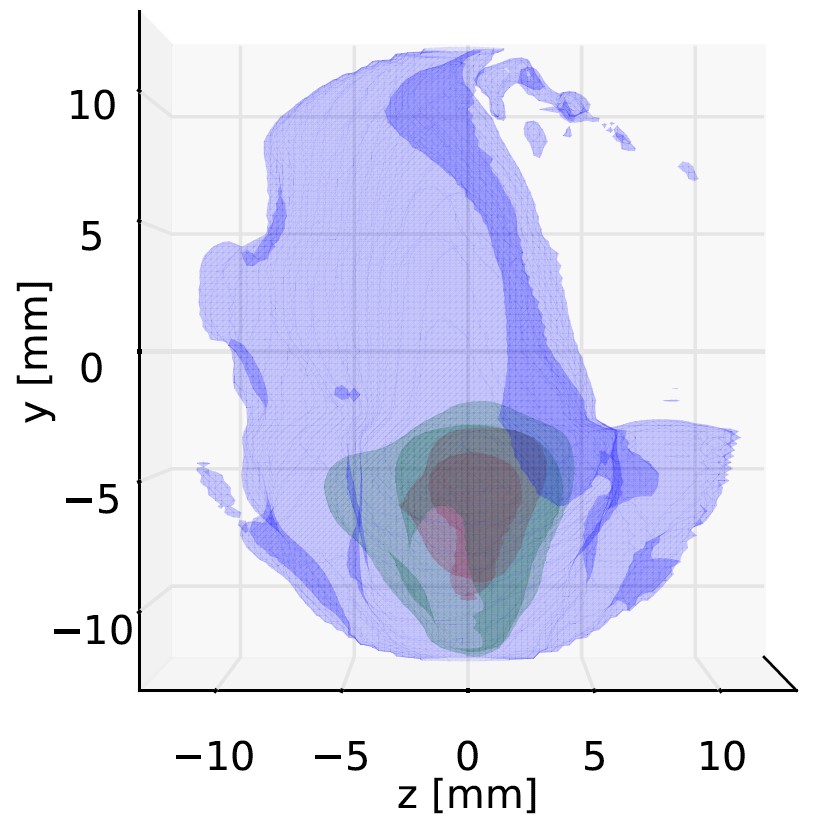}
        \caption{Side}
    \end{subfigure}

    \caption{Reconstructed flame from three poses of the R29 camera.  
        Isocontours shown at 2\% (blue), 40\% (green) and 70\% (red) of the reconstruction's
        maximum value.}
    \label{fig,real,render}
\end{figure}

We placed a burner on a rotating stage and took three images, each separated by
$15^\circ$, with a Raytrix R29 camera attached to a 105 mm 1:2.8D macro lens;
Figure~\ref{fig,real,views} shows the center view.  As
Figure~\ref{fig,real,views} shows, our detector has a region of damaged pixels.
Although one could account for these defects in the reconstruction algorithm by
identifying and down-weighting these pixels via the weights matrix $\W_c$, in
this preliminary experiment we did not perform any such correction.

Some calibration had to be performed to fit our simple plenoptic approach
(Figure~\ref{fig,plenoptic}) to the R29 camera.  More sophisticated calibration
techniques have been proposed by Raytrix~\cite{johannsen:13:otc} and others,
but we use only a very simple method in this work.  We performed a simple
corner-based calibration to determine $D_{\mu m}$, the main lens-microlens
array distance, and used vendor-given values for the main lens focal length,
detector-microlens array distance $D_{\mu d}$, and microlens focal lengths.  We
tuned the aperture of the main lens to produce a simulated checkerboard image
qualitatively similar to the calibration checkerboard image.  The calibrated
parameters themselves are proprietary, but the calibration code we used will be
available under an open source license.

We reconstructed the flame onto a $100^3$ voxel grid with $\GP{\frac{1}{4}}^3$
mm$^3$.  The weighting matrix $\W_c$ was used to extract only the green Bayer
channel from the raw data.  We used a $5 \times 5$ angular discretization
with $N_\text{subset} = 4$ subsets; each iteration took about 50 seconds
and we ran 100 iterations to produce the images in Figure~\ref{fig,real,render}.

Figure~\ref{fig,real,render} shows that the proposed algorithm can successfully
recover the 3D structure of the flame.  There is some stretching in the axial
direction that is particularly noticeable in the lower-valued and cooler (blue)
flame.  More sophisticated calibration and simultaneous capture of multiple
perspectives (instead of the ``rotate-and-shoot'' acquisition used here), may
mitigate these problems.

\section{Conclusions and future work}
\label{sec,conc}

We proposed an efficient light transport-based system model for camera inverse
problems.  The technique models cameras as compositions of light transport
steps that can be efficiently implemented on modern computing hardware.  In
simulated experiments, we modeled a plenoptic 2.0 camera and a traditional
single-lens camera and performed 3D flame reconstructions using a
FISTA-based~\cite{beck:09:afi} algorithm.  We also validated the proposed
algorithm on 3D reconstruction problem with real data taken from a Raytrix R29
camera.

A drawback of many existing camera models for inverse problems is their
reliance on precomputation and sparse linear algebra routines.  These
techniques are mathematically correct but often too inefficient to produce
reconstructions quickly.  The proposed model is efficient and flexible: the
general-purpose reconstruction algorithm in this paper can perform a 3D
reconstruction from three plenoptic camera images in under an hour on a single
GPU, and more specialized algorithms would likely produce faster
reconstructions.  Furthermore, faster algorithms make dynamic time-varying 3D
reconstructions more feasible.

Future work on handling nonideal optics will hopefully improve the
reconstruction results reported in Section~\ref{sec,recon}.  We also plan to
apply the tools in this paper to other lightfield imaging and recovery tasks,
along with more systematic quantitative results and comparisons of camera
configurations for specific applications.

\section*{Acknowledgments}
The authors thank Il Yong Chun for reviewing an early copy of this manuscript
and Christian Heinze of Raytrix for camera specifications.

\bibliographystyle{IEEEtran}

\appendices

\section{Reconstruction cost function Hessian}
\label{sec,hessian}

This appendix shows that the Hessian of the reconstruction cost
function~\eref{eqn,pls} in terms of $\x$ (after minimization over the gains) is
positive semidefinite and easily majorized.

For simplicity, first perform a change of variables to ``absorb'' the weights
$\W_c$ to the system matrix $\A_c$ and the data $\y_c$.
Perform the inner minimization over the camera gains $\GC{\gamma_c}$:
\begin{align}
    \gamma_c &= \frac{\y_c\tr \A_c\x}{\y_c \tr \y_c}.
    \label{eqn,optimal,gain}
\end{align}
Plug~\eref{eqn,optimal,gain} into~\eref{eqn,pls}:
\begin{align}
    \min_{\GC{\gamma_c}} \Psi\GP{\x, \GC{\gamma_c}}
    &=
    \frac{1}{2}\GN{\A_1\x - \y_1}_2^2 + \mathsf{R}\GP{\x}
    \nonumber \\
    & 
    + \sum_{c=2}^{N_c} \frac{1}{2}\GN{%
        \GP{\I - \frac{1}{\GN{\y_c}^2}\y_c\y_c\tr}\A_c \x}_2^2.
\end{align}
The Hessian of this resulting cost function is
\begin{align}
    \hess_\x
    \min_{\GC{\gamma_c}} \Psi\GP{\x, \GC{\gamma_c}}
    &=
    \A_1\tr\A_1 + \hess \R\GP{\x} + 
    \sum_{c=2}^{N_c} \A_c\tr \G_c^2 \A_c
    \nonumber \\
    &\succeq \zeros,
\end{align}
where $\G_c = \I - \frac{1}{\GN{\y_c}^2}\y\y\tr$ is positive semidefinite
but has spectral radius less than unity.  Consequently,
\begin{align}
    \hess_\x
    \min_{\GC{\gamma_c}} \Psi\GP{\x, \GC{\gamma_c}}
    &\preceq
    \sum_{c=1}^{N_c} \A_c\tr\A_c + \hess \R\GP\x
    \nonumber \\
    &\preceq
    \D + \hess\mathsf{R}\GP\x
    \nonumber \\
    &\preceq
    \D + 26 \beta \I,
\end{align}
where $\D$ is the usual diagonal majorizer for the data-fit
terms~\cite{erdogan:99:osa}:
\begin{align}
    \D 
    &=
    \diag_{j}\GC{%
        \GB{%
            \sum_{c=1}^{N_c}
            \A_c\tr\A_c \ones
        }_j
    }.
\end{align}

\end{document}